\documentclass[authoryear,preprint,review,12pt]{elsarticle}

\usepackage{epsfig}

\usepackage{amssymb}
\usepackage{amsthm}

\usepackage{geometry}                
\usepackage[parfill]{parskip}    
\usepackage{amssymb}
\usepackage{comment}
\usepackage{graphicx,caption}
\usepackage{epstopdf}
\usepackage{hyperref}
\usepackage{bigints}
\usepackage{natbib}
\hypersetup{
    colorlinks = true,
    urlcolor   = blue,
    citecolor  = black,
}
\newtheorem{theorem}{Theorem}
\newtheorem{corollary}{Corollary}

\theoremstyle{remark}
\newtheorem{remark}{Remark}

\begin{document}

\begin{frontmatter}



\title{Vorticity Alignment with Lyapunov Vectors and Rate-of-Strain Eigenvectors}

\author[1]{A. Encinas-Bartos}

\author[1]{G. Haller \corref{cor1}}
\ead{georgehaller@ethz.ch}

 \cortext[cor1]{Corresponding author}

 \affiliation[1]{organization={Institute for Mechanical Systems, ETH Zurich}, 
                 city={Zurich}, 
                 country={Switzerland}}


\begin{abstract}
We derive asymptotic estimates for the projection of the vorticity onto principal directions of material stretching in 3D flows. In flows with pointwise bounded vorticity, these estimates predict vorticity alignment with Lyapunov vectors along trajectories with positive Lyapunov exponents. Specifically, we find that in inviscid flows with conservative body forces, the vorticity exactly aligns with the intersection of the planes orthogonal to the dominant forward and backward Lyapunov vectors along trajectories with positive Lyapunov exponent. Furthermore, we derive asymptotic estimates for the vorticity alignment with the intermediate eigenvector of the rate of strain tensor for viscous flows under general forcing. We illustrate these results on explicit solutions of Euler’s equation and on direct numerical simulations of homogeneous isotropic turbulence.
\end{abstract}



\begin{keyword}
Vorticity alignment, Arnold-Beltrami-Childress flow, Homogenous Isotropic Turbulence, Dynamical Systems


\end{keyword}

\end{frontmatter}


\section{\label{sec:Intro}Introduction}
\label{sec: Intro}
Investigating the distribution of the vorticity in three dimensional (3D) flows is crucial for uncovering the intricate and chaotic dynamics driving the formation of the complex structures in turbulence. Specifically, understanding the stretching and tilting of the vorticity along a Lagrangian particle trajectory sheds light on the underlying mechanisms driving the formation and evolution of these structures.  \\ \\
The geometry of vortex lines is closely linked to the global existence of solutions in 3D Navier stokes flows \citep{Constantin1993, Constantin1994, Deng2005, Constantin2020} and to the vortex stretching mechanism \citep{Galanti1997, Buaria2020a, Buaria2020b,Carbone2020}. Pioneering experiments from \citet{Taylor1938} showed that vortex stretching is responsible for the vorticity production and dissipation in isotropic turbulence. In the absence of external forcing, the dynamics of the vorticity $ \boldsymbol{\omega}(\boldsymbol{x}, t) \in \mathbb{R}^3 $ at a point $ \boldsymbol{x} \in \mathbb{R}^3 $ and time $ t \in \mathbb{R} $ depend on both the rate-of-strain tensor, $ \mathbf{S}(\boldsymbol{x}, t)=\dfrac{1}{2}(\boldsymbol{\nabla} \boldsymbol{v}(\boldsymbol{x},t)^T+\boldsymbol{\nabla}\boldsymbol{v}(\boldsymbol{x},t)) $ of the velocity field $ \boldsymbol{v}(\boldsymbol{x},t) $ and on the viscous term $ \nu \boldsymbol{\Delta \omega}(\boldsymbol{x}, t) $, as seen from the vorticity transport equation
\begin{equation}
\dfrac{D}{Dt}\boldsymbol{\omega}(\boldsymbol{x}, t) = \mathbf{S}(\boldsymbol{x}, t)\boldsymbol{\omega}(\boldsymbol{x}, t) + \nu \boldsymbol{\Delta \omega}(\boldsymbol{x}, t).
\end{equation}
This equation shows that in Euler flows, i.e. inviscid flows with $ \nu = 0 $, vortex lines evolve as material lines. Similarly, in high-Reynolds number flows, the non-local viscous contribution in the Navier-Stokes equation is negligible over short time-scales and the vorticity evolves approximately as a material vector \citep{Taylor1938}. The instantaneous stretching of the vorticity vector $ \boldsymbol{\omega} $ is therefore closely related to the eigenvectors and eigenvalues of the rate-of-strain tensor $ \mathbf{S}$. In inviscid flows, the long-term stretching and tilting of the vorticity vector along a Lagrangian particle trajectory is most naturally described in a Lagrangian setting through the eigenvalues and eigenvectors of the Cauchy-Green strain tensor \citep{Ni2014}. \\ \\
Over the last two decades, extensive research has been dedicated to investigate statistical correlations between the vorticity and the rate-of-strain tensor. Experimental \citep{Tsinober1992, Kit1993} and numerical \citep{Ashurst1987, She1991, Jimenez1992, Huang1996} studies in turbulence have suggested that the vorticity tends to align preferentially  with the intermediate eigenvector of $ \mathbf{S}(\boldsymbol{x},t) $, whose corresponding eigenvalue tends to be positive. Vortex lines in regions of high vorticity tend to be straight and well-aligned, thereby displaying highly organized structures \citep{Galanti1996}. Regions of high vorticity frequently arise in connection with a preferential alignment with the intermediate eigenvector of the rate of strain tensor \citep{Ashurst1987, Jimenez1992, Jimenez1993}. Other studies by \citet{Kerr1985} and \citet{Jimenez1992} proposed that, in regions of approximately 2D velocity, the vorticity aligns with the intermediate rate-of-strain eigenvector. In that 2D case, the vorticity is neither stretched nor tilted. Rather, the alignment occurs because the vorticity and the intermediate eigenvector of $ \mathbf{S}(\boldsymbol{x},t) $ are approximately orthogonal to the plane of the flow in such regions. \\ \\
In addition to the studies that focus on the instantaneous alignment between the vorticity and the eigenvectors of the rate of strain tensor,  \citet{Xu2011} and \citet{Pumir2013} started incorporating dynamical information into the analysis. They considered the evolution of the vorticity along a trajectory and compared the alignment between the vorticity at time $ t +\Delta t $ with the most stretching direction specified by the strain rate at time $ t $. They found that over short times, the vorticity evolving along a trajectory aligns with the most stretching direction of $ \mathbf{S}(\boldsymbol{x},t) $. However, for $ \Delta t > 2 \tau_{\eta} $, where $ \tau_{\eta} $ is the Kolmogorov time-scale, the observed alignment breaks down. This is in agreement with the theory of Objective Eulerian Coherent Structures (OECS) developed by \citet{Serra2016}: over short times, material elements align with the dominant eigenvector of $ \mathbf{S}(\boldsymbol{x},t) $. This alignment persists along a trajectory as long as the eigenvectors of the rate of strain tensor vary slowly along that trajectory \citep{Dresselhaus1992, Tabor1994}. This persistence of the rate of strain hypothesis was first proposed by \citet{Batchelor1952, Batchelor1956} as a convenient assumption that allows the study of deformation of material elements under a fixed strain rate. Numerical and experimental Lagrangian data was not readily available at that time and hence this assumption could not be verified until the 1980s. Numerical simulations performed by Girimaji and Pope \citep{Girimaji1990, Pope1990} finally showed, however, that the persistence of strain assumption remains valid only over times of the order of $ 2\tau_{\eta} $. \\ \\
As an alternative, \citet{Hamlington2008b, Hamlington2008a} proposed to split the rate-of-strain tensor into a local and non-local component  $ \mathbf{S} = \mathbf{S}^L+\mathbf{S}^B $. Here $ \mathbf{S}^L $ is the local rate of strain induced by a vortical structure in a spherical neighborhood, and $\mathbf{S}^B$ is the non-local background rate of strain induced in the vicinity of the structure by all the remaining vorticity. This decomposition is non-unique since it depends on the radius of the local neighborhood. Furthermore, the non-local rate of strain $ \mathbf{S}^B $ is not easily computable in practice since it relies on higher-order spatial derivatives of the velocity field. \citet{Hamlington2008b, Hamlington2008a} showed through numerical simulations that the vorticity preferentially aligns both with the most stretching eigenvector of the non-local (background) rate of strain $ \mathbf{S}^B $ and with the intermediate eigenvector of the regular rate of strain $ \mathbf{S} $. \\ \\
The evolution of the vorticity can also be studied in a Lagrangian setting, i.e., along a trajectory. \citet{Ni2014} compared numerically the evolution of the vorticity with that of thin rods (which approximate material fibers) dispersed in a fluid. Both the vorticity and the rods were found to align with the most stretching direction of the Cauchy-Green strain tensor under forward advection. The alignment of the vorticity was argued to be weaker compared to that of the thin rods, since the viscous term was expected to cause its dynamics to divert from purely material evolution. We will further discuss the validity of these findings in view of the results of this paper (see the discussion after Remark 1). \\ \\
Finally in further related work, \citet{Tsinober2000, Guala2005, Luthi2005} studied the evolution of the vorticity along a Lagrangian particle trajectory with respect to the rate of strain eigenvectors for non-zero viscosity. They showed numerically that over longer time intervals, the viscous term enhances the preferential alignment of the vorticity with the intermediate eigenvector of the rate-of-strain tensor.\\ \\
Here we derive precise estimates on the vorticity evolution in a Lagrangian frame that imply various alignment results for incompressible flows with bounded vorticity. First, for inviscid flows with potential body forces, we obtain an exact alignment between the vorticity and the normal of the plane of the forward and backward Lyapunov vectors (i.e. asymptotic limits of the forward and backward material stretching exponents). This normal coincides with the intermediate Lyapunov vector along trajectories whose forward and backward dynamics are identical. We similarly provide estimates for the alignment of the vorticity with principal stretching directions of the rate of strain tensor in general forced viscous flows. These estimates depend on the weighted average between the exponential of the principal strains and the variation of the rate of strain tensor along a trajectory times its vorticity plus the curl of the external forces (viscous and non-potential). We illustrate all these results on laminar and turbulent flow data.

\section{Lagrangian setup}
Consider a 3D incompressible velocity field 
\begin{equation}
\dot{\boldsymbol{x}}=\boldsymbol{v}(\boldsymbol{x},t),\quad\boldsymbol{\nabla}\cdot\boldsymbol{v}=0,\label{eq:velocity field}
\end{equation} on the spatial domain $\boldsymbol{x}\in U\subset\mathbb{R}^{3}$. A trajectory of this velocity field that starts from the position $\boldsymbol{x}_{0}$ at time $t_{0}$ is denoted by $\boldsymbol{x}(t;t_{0},\boldsymbol{x}_{0})$. The flow map $\mathbf{F}_{t_{0}}^{t}\colon\boldsymbol{x}_{0}\mapsto\boldsymbol{x}(t;t_{0},\boldsymbol{x}_{0})$ is then defined as the mapping from initial positions into current positions along trajectories. The equation of variations (or linearized flow) along $\boldsymbol{x}(t;t_{0},\boldsymbol{x}_{0})$ is a homogeneous system of non-autonomous linear ordinary differential equations (ODEs) of the form 
\begin{equation}
\dot{\boldsymbol{\xi}}(t)=\left[\boldsymbol{\nabla}\boldsymbol{v}\left(\boldsymbol{x}(t;t_{0},\boldsymbol{x}_{0}),t\right)\right]\boldsymbol{\xi}(t).\label{eq:equation of variations}
\end{equation}
This system of ODEs describes the material evolution of an infinitesimal perturbation $\boldsymbol{\xi}(t)$ along $\boldsymbol{x}(t;t_{0},\boldsymbol{x}_{0})$, starting from an initial perturbation $\boldsymbol{\xi}(t_{0})=\boldsymbol{\xi}_{0}$. The general solution of this system is
\begin{equation}
\boldsymbol{\xi}(t)=\boldsymbol{\nabla}\mathbf{F}_{t_{0}}^{t}\left(\boldsymbol{x}_{0}\right)\boldsymbol{\xi}_{0}.\label{eq:solution of equation of variations-1}
\end{equation} 
Taking the inner product of (\ref{eq:equation of variations}) with $\boldsymbol{\xi}(t)$ and integrating in time gives that the length of $\boldsymbol{\xi}(t)$ evolves as \[\left|\boldsymbol{\xi}(t)\right|^{2}=\left\langle \boldsymbol{\xi}_{0},\mathbf{C}_{t_{0}}^{t}\left(\boldsymbol{x}_{0}\right)\boldsymbol{\xi}_{0}\right\rangle ,\qquad\mathbf{C}_{t_{0}}^{t}\left(\boldsymbol{x}_{0}\right)=\left[\boldsymbol{\nabla}\mathbf{F}_{t_{0}}^{t}\left(\boldsymbol{x}_{0}\right)\right]^{\mathrm{T}}\boldsymbol{\nabla}\mathbf{F}_{t_{0}}^{t}\left(\boldsymbol{x}_{0}\right), \] with $\mathbf{C}_{t_{0}}^{t}\left(\boldsymbol{x}_{0}\right)$ denoting the right Cauchy\textendash Green strain tensor. This symmetric, positive definite tensor has three real, positive eigenvalues that can be ordered as 
\begin{align}
0<\lambda_{1}(t,t_{0},\boldsymbol{x}_{0})\leq\lambda_{2}(t,t_{0},\boldsymbol{x}_{0})\leq\lambda_{3}(t,t_{0},\boldsymbol{x}_{0}),\notag \\
\lambda_{1}(t;t_{0},\boldsymbol{x}_{0})\lambda_{2}(t;t_{0},\boldsymbol{x}_{0})\lambda_{3}(t;t_{0},\boldsymbol{x}_{0})=1,\label{eq:lambda_i relations}
\end{align} where the second relation follows from the incompressibility of $\boldsymbol{v}(\boldsymbol{x},t)$. The corresponding unit eigenvectors of $\mathbf{C}_{t_{0}}^{t}\left(\boldsymbol{x}_{0}\right)$ satisfy
\begin{align} 
\mathbf{C}_{t_{0}}^{t}\left(\boldsymbol{x}_{0}\right)\boldsymbol{\xi}_{j}(t;t_{0},\boldsymbol{x}_{0})=\lambda_{j}(t;t_{0},\boldsymbol{x}_{0})\boldsymbol{\xi}_{j}(t;t_{0},\boldsymbol{x}_{0}),\\
\left\langle \boldsymbol{\xi}_{i},
\boldsymbol{\xi}_{j}\right\rangle=\delta_{ij},\quad i,j=1,2,3,\quad i\neq j.\end{align}
The vectors $\boldsymbol{\xi}_{j}(t;t_{0},\boldsymbol{x}_{0})$ are also the right singular vectors of $\boldsymbol{\nabla}\mathbf{F}_{t_{0}}^{t}\left(\boldsymbol{x}_{0}\right)$.
A useful identity is 
\begin{equation}
\boldsymbol{\nabla}\mathbf{F}_{t_{0}}^{t}\left(\boldsymbol{x}_{0}\right)\boldsymbol{\xi}_{j}(t;t_{0},\boldsymbol{x}_{0})=\sqrt{\lambda_{j}(t;t_{0},\boldsymbol{x}_{0})}\boldsymbol{\eta}_{j}(t;t_{0},\boldsymbol{x}_{0}),\quad j=1,2,3,\label{eq:left-right singular vectors of deformaiton gradient}
\end{equation} where the orthonormal vectors $\boldsymbol{\eta}_{j}(t;t_{0},\boldsymbol{x}_{0})$ are the left singular vectors of $\boldsymbol{\nabla}\mathbf{F}_{t_{0}}^{t}\left(\boldsymbol{x}_{0}\right)$ corresponding to the singular values $\sqrt{\lambda_{j}(t;t_{0},\boldsymbol{x}_{0})}$, respectively (see, e.g., \citet{Haller2023}). Based on formula (\ref{eq:left-right singular vectors of deformaiton gradient}), the finite-time Lyapunov exponent (FTLE) is often used to identify an overall growth exponent in the temporal evolution of $\lambda_{3}(t,t_{0},\boldsymbol{x}_{0})$ over a time interval $[t_{0},t_{1}]$ via the formula 
\begin{equation}
\mathrm{FTLE}_{t_{0}}^{t_{1}}\left(\boldsymbol{x}_{0}\right)=\frac{1}{\left|t_{1}-t_{0}\right|}\log\sqrt{\lambda_{3}(t_{1};t_{0},\boldsymbol{x}_{0})}.\label{eq:FTLEdef}
\end{equation} A well-defined asymptotic growth exponent for $\sqrt{\lambda_{3}(t_{1};t_{0},\boldsymbol{x}_{0})}$
may continue to exist as $t_{1}\to\pm\infty$. In such cases, the maximal forward- and backward Lyapunov exponents $\mu_{3}^{\pm}\left(t_{0},\boldsymbol{x}_{0}\right)$ of the trajectory $\boldsymbol{x}(t;t_{0},\boldsymbol{x}_{0})$ can be defined as the limiting singular values
\begin{equation}
\mu_{3}^{\pm}\left(\boldsymbol{x}_{0},t_{0}\right)=\lim_{t\to\pm\infty}\frac{1}{\left|t-t_{0}\right|}\log\sqrt{\lambda_{3}(t,t_{0},\boldsymbol{x}_{0})}.\label{eq:Lyapuov exponent}
\end{equation}
Similarly, the dominant forward and backward Lyapunov vectors can be defined as the limiting dominant singular vectors:
\begin{equation}
\boldsymbol{\zeta}_{3}^{\pm}(\boldsymbol{x}_{0},t_{0})=\lim_{t\to\pm\infty}\boldsymbol{\xi}_{3}(t;t_{0},\boldsymbol{x}_{0}).\label{eq:Lyapunov vector}
\end{equation}
The existence of the limits (\ref{eq:Lyapuov exponent}) and (\ref{eq:Lyapunov vector}) cannot be \emph{a priori} guaranteed for trajectories of a general velocity field (\ref{eq:velocity field}). If, however, the incompressible velocity field $\boldsymbol{v}(\boldsymbol{x},t)\equiv\boldsymbol{v}(\boldsymbol{x})$ is steady and the domain $U$ is compact and invariant under the flow, then the limits (\ref{eq:Lyapuov exponent}) and (\ref{eq:Lyapunov vector}) (along with further Lyapunov vectors and exponents corresponding to the remaining singular values of $\boldsymbol{\nabla}\mathbf{F}_{t_{0}}^{t}\left(\boldsymbol{x}_{0}\right)$) exist for almost all $\boldsymbol{x}_{0}$ by the multiplicative ergodic theorem of \citet{Oseledets1968}. This theorem is originally stated for
maps, which in turn implies the existence of the limits (\ref{eq:Lyapuov exponent}) and (\ref{eq:Lyapunov vector}) for time-periodic flows as well via the Poincaré maps associated with those flows. We note that alternative
definitions for Lyapunov vectors are also available and generally give different results (see, e.g.,  \citet{Wolfe2007}).
\section{Eulerian setup}
We assume now that in an inertial frame, the velocity field $\boldsymbol{v}(\boldsymbol{x},t)$ in Eq.(\ref{eq:velocity field}) is a solution of the 3D, incompressible Navier\textendash Stokes equation
\begin{equation}
\frac{D\boldsymbol{v}(\boldsymbol{x},t)}{Dt}=-\frac{1}{\rho}\nabla p(\boldsymbol{x},t)+\nu\Delta\boldsymbol{v}(\boldsymbol{x},t)+\boldsymbol{f}(\boldsymbol{x},t),\label{eq:Navier--Stokes}
\end{equation}
with the constant density $\rho$, the pressure field $p(\boldsymbol{x},t)$
and the body force vector $\boldsymbol{f}(\boldsymbol{x},t).$ Taking the curl of this equation gives the incompressible vorticity transport equation
\begin{equation}
\frac{D\boldsymbol{\omega}}{Dt}=\left[\boldsymbol{\nabla}\boldsymbol{v}\right]\boldsymbol{\omega}+\nu\Delta\boldsymbol{\omega}+\boldsymbol{\nabla}\times\boldsymbol{f},\label{eq:vorticity transport}
\end{equation}
 for the vorticity $\boldsymbol{\omega}=\boldsymbol{\nabla}\times\boldsymbol{v}$. We note that the last term on the right\textendash hand side of Eq.(\ref{eq:vorticity transport}) vanishes if all body forces acting on the fluid are potential forces. \\ \\
Analogously to the maximal material stretching exponent quantified by the $\mathrm{FTLE}_{t_{0}}^{t_{1}}\left(\boldsymbol{x}_{0}\right)$ field defined in (\ref{eq:FTLEdef}), we can define the vorticity
stretching exponent (VSE) field as 
\begin{equation}
\mathrm{VSE}_{t_{0}}^{t_{1}}\left(\boldsymbol{x}_{0}\right)=\frac{1}{\left|t_{1}-t_{0}\right|}\log\frac{\left|\boldsymbol{\omega}\left(\boldsymbol{x}(t_{1};t_{0},\boldsymbol{x}_{0}),t_1\right)\right|}{\left|\boldsymbol{\omega}\left(\boldsymbol{x}_{0},t_{0}\right)\right|},\label{eq:VSEdef}
\end{equation}  which we will use in our later formulas and numerical experiments to characterize the long\textendash term evolution of the vorticity magnitude. Vorticity alignment is often studied with respect to the eigenvectors
$\boldsymbol{e}_{j}(\boldsymbol{x},t)$ of the rate\textendash of\textendash strain tensor \[ \mathbf{S}(\boldsymbol{x},t)=\frac{1}{2}\left[\left[\boldsymbol{\nabla}\boldsymbol{v}\left(\boldsymbol{x},t\right) \right]^{\mathrm{T}}+\boldsymbol{\nabla}\boldsymbol{v}\left(\boldsymbol{x},t\right)\right].\] This symmetric, positive semidefinite tensor is objective, i.e., it transforms properly as a linear operator under any Euclidean frame change of the form $\boldsymbol{x}=\mathbf{Q}(t)\boldsymbol{y}+\boldsymbol{b}(t)$, where $\mathbf{Q}(t)\in\mathrm{SO}(3)$ is an arbitrary time\textendash dependent rotation tensor and $\boldsymbol{b}(t)\in\mathbb{R}^{3}$ is an arbitrary translation vector. The eigenvectors $\boldsymbol{e}_{j}(\boldsymbol{x},t)$ of $\mathbf{S}(\boldsymbol{x},t)$ are also objective, because the strain eigenvectors $\tilde{\boldsymbol{e}}_{j}(\boldsymbol{y},t)$ in the $\boldsymbol{y}$-frame come out to be
\begin{equation}
\tilde{\boldsymbol{e}}_{j}(\boldsymbol{y},t)=\mathbf{Q}^{\mathrm{T}}(t)\boldsymbol{e}_{j}(\boldsymbol{x},t).\label{eq:strain eigenvector transformation}
\end{equation} In contrast, the vorticity $\tilde{\boldsymbol{\omega}}$ of the transformed velocity field obeys the formula
\begin{equation}
\tilde{\boldsymbol{\omega}}(\boldsymbol{y},t)=\mathbf{Q}^{\mathrm{T}}(t)\left[\boldsymbol{\omega}(\boldsymbol{x},t)-\dot{\boldsymbol{q}}(t)\right],\label{eq:vorticity transformation}
\end{equation}
 where $\dot{\boldsymbol{q}}(t)$ is the vorticity associated with the frame change (see \citet{Haller2023}). This shows that unlike $\boldsymbol{e}_{j}(\boldsymbol{x},t)$, the vorticity field $\boldsymbol{\omega}(\boldsymbol{x},t)$ is not objective.   A comparison of formulas (\ref{eq:strain eigenvector transformation}) and (\ref{eq:vorticity transformation}) implies that any possible alignment (or lack thereof) observed between the vorticity and the rate\textendash of\textendash strain eigenvectors is inherently frame\textendash dependent due to the
non\textendash preservation of the inner product under a general frame change:
\begin{align*}
\left\langle \tilde{\boldsymbol{\omega}}(\boldsymbol{y},t),\tilde{\mathbf{e}}_{j}(\boldsymbol{y},t)\right\rangle  & =\left\langle \boldsymbol{\omega}(\boldsymbol{x},t),\boldsymbol{e}_{j}(\boldsymbol{x},t)\right\rangle +\left\langle \dot{\boldsymbol{q}}(t),\boldsymbol{e}_{j}(\boldsymbol{x},t)\right\rangle .
\end{align*}
This is also true for our upcoming alignment results, which will specifically be valid in inertial frames. In such frames, the fundamental form of Eq.(\ref{eq:Navier--Stokes}) remains unchanged. 
\section{Vorticity alignment for Euler flows in the Lagrangian frame}
\label{sec:Main-results}
Under potential body forces, the inviscid limit of Eq.(\ref{eq:vorticity transport}),  restricted to a trajectory $\boldsymbol{x}(t;t_{0},\boldsymbol{x}_{0})$ of the velocity field $\boldsymbol{v}(\boldsymbol{x},t)$, is a non\textendash autonomous, homogeneous linear system of ODEs of the form 
\begin{equation}
\frac{D\boldsymbol{\omega}}{Dt}=\left[\boldsymbol{\nabla}\boldsymbol{v}\left(\boldsymbol{x}(t;t_{0},\boldsymbol{x}_{0}),t\right)\right]\boldsymbol{\omega}.\label{eq:inviscid vorticity tranpsort eq}
\end{equation}
This linear system formally coincides with the equation of variations, and hence by Eq.(\ref{eq:solution of equation of variations-1}), its solution can be written as
\begin{equation}
\centering
\boldsymbol{\omega}\left(\boldsymbol{x}(t;t_{0},\boldsymbol{x}_{0}),t\right)=\boldsymbol{\nabla}\mathbf{F}_{t_{0}}^{t}\left(\boldsymbol{x}_{0}\right)\boldsymbol{\omega}_{0}\left(\boldsymbol{x}_{0}\right).\label{eq:omega(t)}
\end{equation}
Using the shorthand notation
\[\boldsymbol{\omega}_{t}(\boldsymbol{x}_{0}):=\boldsymbol{\omega}\left(\boldsymbol{x}(t;t_{0},\boldsymbol{x}_{0}),t\right), \]
we square both sides of Eq.(\ref{eq:omega(t)}) to obtain
\begin{align}
\left|\boldsymbol{\omega}_{t}(\boldsymbol{x}_{0})\right|^{2} & =\left|\boldsymbol{\nabla}\mathbf{F}_{t_{0}}^{t}\left(\boldsymbol{x}_{0}\right)\boldsymbol{\omega}_{t_{0}}\left(\boldsymbol{x}_{0}\right)\right|^{2}, \label{eq:main estimate}
\end{align} whose right-hand side can be upper estimated as
\begin{align} \left|\boldsymbol{\nabla}\mathbf{F}_{t_{0}}^{t}\left(\boldsymbol{x}_{0}\right)\boldsymbol{\omega}_{t_{0}}\left(\boldsymbol{x}_{0}\right)\right|^{2}&=\left\langle \boldsymbol{\omega}_{t_{0}}\left(\boldsymbol{x}_{0}\right),\mathbf{C}_{t_{0}}^{t}\left(\boldsymbol{x}_{0}\right)\boldsymbol{\omega}_{t_{0}}\left(\boldsymbol{x}_{0}\right)\right\rangle \\
&\leq\lambda_{3}(t;t_{0},\boldsymbol{x}_{0})\left|\boldsymbol{\omega}_{t_{0}}\left(\boldsymbol{x}_{0}\right)\right|^{2}.
\end{align}
At the same time, using the identity (\ref{eq:left-right singular vectors of deformaiton gradient}),
we obtain the lower estimate
\begin{align}
\left|\boldsymbol{\nabla}\mathbf{F}_{t_{0}}^{t}\left(\boldsymbol{x}_{0}\right)\boldsymbol{\omega}_{t_{0}}\left(\boldsymbol{x}_{0}\right)\right|^{2} & =\left|\boldsymbol{\nabla}\mathbf{F}_{t_{0}}^{t}\left(\boldsymbol{x}_{0}\right)\sum_{j=1}^{3}\left\langle \boldsymbol{\omega}_{t_{0}}\left(\boldsymbol{x}_{0}\right),\boldsymbol{\xi}_{j}(t;t_{0},\boldsymbol{x}_{0})\right\rangle \right. \nonumber \\
&\left. \cdot \boldsymbol{\xi}_{j}(t;t_{0},\boldsymbol{x}_{0})\right|^{2}\nonumber \\
 &=\left|\sum_{j=1}^{3}\left\langle \boldsymbol{\omega}_{t_{0}}\left(\boldsymbol{x}_{0}\right),\boldsymbol{\xi}_{j}(t;t_{0},\boldsymbol{x}_{0})\right\rangle \right. \nonumber \\
 &\left. \cdot \sqrt{\lambda_{j}(t;t_{0},\boldsymbol{x}_{0})}\boldsymbol{\eta}_{j}(t;t_{0},\boldsymbol{x}_{0})\right|^{2}\nonumber \\
 & =\sum_{j=1}^{3}\left\langle \boldsymbol{\omega}_{t_{0}}\left(\boldsymbol{x}_{0}\right),\boldsymbol{\xi}_{j}(t;t_{0},\boldsymbol{x}_{0})\right\rangle ^{2}\lambda_{j}(t;t_{0},\boldsymbol{x}_{0})\nonumber \\
 & \geq\left\langle \boldsymbol{\omega}_{t_{0}}\left(\boldsymbol{x}_{0}\right),\boldsymbol{\xi}_{j}(t;t_{0},\boldsymbol{x}_{0})\right\rangle ^{2}\lambda_{j}(t;t_{0},\boldsymbol{x}_{0}), \label{eq:identity 3}\\
& j=1,2,3. \nonumber
\end{align} 
Combining the expressions (\ref{eq:main estimate}) and (\ref{eq:identity 3}) with $j=3$, we obtain the estimate
\begin{align}
\left\langle \boldsymbol{\omega}_{t_{0}}\left(\boldsymbol{x}_{0}\right),\boldsymbol{\xi}_{3}(t;t_{0},\boldsymbol{x}_{0})\right\rangle ^{2}\lambda_{3}(t;t_{0},\boldsymbol{x}_{0}) \leq \left|\boldsymbol{\omega}_{t}(\boldsymbol{x}_{0})\right|^{2},
\end{align} or, equivalently, 
\begin{align}
\left\langle \boldsymbol{\omega}_{t_{0}}\left(\boldsymbol{x}_{0}\right),\boldsymbol{\xi}_{3}(t;t_{0},\boldsymbol{x}_{0})\right\rangle ^{2}\leq\frac{\left|\boldsymbol{\omega}_{t}(\boldsymbol{x}_{0})\right|^{2}}{\lambda_{3}(t;t_{0},\boldsymbol{x}_{0})}. \label{eq:first omega0 estimate}
\end{align}
Using the unit vector
\[
\mathbf{e}_{\boldsymbol{\omega}_{0}}\left(\boldsymbol{x}_{0}\right)=\frac{\boldsymbol{\omega}_{t_{0}}\left(\boldsymbol{x}_{0}\right)}{\left|\boldsymbol{\omega}_{t_{0}}\left(\boldsymbol{x}_{0}\right)\right|},
\] along with Eq.(\ref{eq:first omega0 estimate}), we obtain the upper estimate
\begin{align}
\left\langle \mathbf{e}_{\boldsymbol{\omega}_{0}}\left(\boldsymbol{x}_{0}\right),\boldsymbol{\xi}_{3}(t;t_{0},\boldsymbol{x}_{0})\right\rangle ^{2}&\leq \frac{1}{\lambda_{3}(t;t_{0},\boldsymbol{x}_{0})} \frac{\left|\boldsymbol{\omega}_{t}\left(\boldsymbol{x}_{0}\right)\right|^{2}}{\left|\boldsymbol{\omega}_{t_{0}}\left(\boldsymbol{x}_{0}\right)\right|^{2}},
\end{align} which can be rewritten as 
\begin{equation}
\left|\left\langle \mathbf{e}_{\boldsymbol{\omega}_{0}}\left(\boldsymbol{x}_{0}\right),\boldsymbol{\xi}_{3}(t;t_{0},\boldsymbol{x}_{0})\right\rangle \right|\leq\mathcal{B}_{t_{0}}^{t}\left(\boldsymbol{x}_{0}\right):=e^{\left[\mathrm{VSE}_{t_{0}}^{t}\left(\boldsymbol{x}_{0}\right)-\mathrm{FTLE}_{t_{0}}^{t}\left(\boldsymbol{x}_{0}\right)\right]\left(t-t_{0}\right)}.\label{eq: cal B}
\end{equation}
This last inequality suggests that in incompressible Euler flows, material stretching drives the vorticity vector away from aligning with $\boldsymbol{\xi}_{3}$, whereas vorticity stretching has the opposite effect. \\ \\
Assume now that the largest forward and backward Lyapunov exponents $\mu_{3}^{\pm}\left(t_{0},\boldsymbol{x}_{0}\right)$ along the trajectory
$ \boldsymbol{x}\left(t;t_{0},\boldsymbol{x}_{0}\right)$ exist and are positive, i.e., 
\begin{equation} 
0<\mu_{3}^{\pm}\left(\boldsymbol{x}_{0},t_0\right):=\lim_{t\to\pm\infty}\frac{1}{2\left(t-t_{0}\right)}\log\lambda_{3}(t;t_{0},\boldsymbol{x}_{0})<\infty,  \label{eq: LyapunovExponent}
\end{equation}  which implies that
\begin{align}
\lim_{t \rightarrow \pm \infty} \mathrm{FTLE}_{t_0}^t(\boldsymbol{x}_0) &= \mu^{\pm}_{3}(\boldsymbol{x}_0),\\
\lim_{t \rightarrow \pm \infty} \boldsymbol{\xi}_3(t;t_0,\boldsymbol{x}_0) &= \boldsymbol{\zeta}_3^{\pm}(\boldsymbol{x}_0,t_0) .
\end{align}
For flows with pointwise uniformly bounded vorticity, the first term on the right-hand side of the inequality (\ref{eq: cal B}) vanishes in the limits $t\to\pm\infty$. Using the asymptotic estimates we then deduce the following alignment results for incompressible Euler flows defined on a compact invariant domain $U$ subject to potential body forces.
\begin{theorem}
\label{thm: 1}Consider an incompressible Euler velocity field $\boldsymbol{v}(\boldsymbol{x},t)$, defined on the invariant flow region $U$ with pointwise uniformly bounded vorticity, subject only to potential forces. Assume that along a trajectory $\boldsymbol{x}(t;t_{0},\boldsymbol{x}_{0}),$ the maximal Lyapunov exponents $\mu_{3}^{\pm}\left(\boldsymbol{x}_{0},t_{0}\right)$
are well defined and positive, and the corresponding dominant Lyapunov vectors $\boldsymbol{\zeta}_{3}^{\pm}(\boldsymbol{x}_{0},t_{0})$ satisfy the non-degeneracy condition 
\begin{equation}
\boldsymbol{\zeta}_{3}^{+}(\boldsymbol{x}_{0},t_{0})\nparallel\boldsymbol{\zeta}_{3}^{-}(\boldsymbol{x}_{0},t_{0}).\nonumber
\end{equation}
Define then \[ \boldsymbol{n}(\boldsymbol{x}_0,t_0) = \boldsymbol{\zeta}_{3}^{+}(\boldsymbol{x}_{0},t_{0})\times\boldsymbol{\zeta}_{3}^{-}(\boldsymbol{x}_{0},t_{0}) \] as the vector along the intersection of the two planes orthogonal to $ \boldsymbol{\zeta}_3^{\pm}(\boldsymbol{x}_0) $. \\
Then, the vorticity $\boldsymbol{\omega}\left(\boldsymbol{x}_{0},t_{0}\right)$ exactly aligns with $ \boldsymbol{n}(\boldsymbol{x}_0,t_0) $, i.e.,
\begin{equation}
\boldsymbol{\omega}\left(\boldsymbol{x}_{0},t_{0}\right) \parallel  \boldsymbol{n}(\boldsymbol{x}_0,t_0).
\end{equation}
\end{theorem}
Theorem \ref{thm: 1} states that in order for the vorticity to remain bounded along a trajectory with positive Lyapunov exponents, it must be orthogonal to the most stretching directions in both forward and backward time. This direction is well defined as long as the forward and backward most stretching directions are distinct.
If, in addition, the flow domain is compact and the velocity field is steady or time-periodic, then the Lyapunov vectors and exponents are guaranteed to exist for almost all initial conditions $ \boldsymbol{x}_0 $ according to the multiplicative ergodic theory of \citet{Oseledets1968}. We, therefore, obtain the following strengthened result in that case.
\begin{corollary}
\label{cor: 1}
In a steady or time-periodic, incompressible Euler velocity field $\boldsymbol{v}(\boldsymbol{x},t)$,
defined on a compact invariant domain $U$ with pointwise uniformly bounded vorticity subject only to potential body forces, we have, at almost all initial points $\boldsymbol{x}_{0}$, the exact alignment \[
\boldsymbol{\omega}(\boldsymbol{x}_{0})\parallel\boldsymbol{n}(\boldsymbol{x}_0).
\]
\end{corollary}
\begin{remark}
We now discuss a special setting in which the vorticity $\boldsymbol{\omega}(\boldsymbol{x}_{0},t_{0})$
specifically aligns with the asymptotic limit of the intermediate left Cauchy-Green eigenvector, $\boldsymbol{\xi}_{2}(t;t_{0},\boldsymbol{x}_{0})$.
We first note that a total of three pairs of Lyapunov exponents and vectors can be defined via the limits
\begin{align}
\mu_{j}^{\pm}\left(\boldsymbol{x}_{0},t_{0}\right)=\lim_{t\to\pm\infty}\frac{1}{\left|t-t_{0}\right|}\log\sqrt{\lambda_{j}(t,t_{0},\boldsymbol{x}_{0})}, \\\boldsymbol{\zeta}_{j}^{\pm}(\boldsymbol{x}_{0},t_{0})=\lim_{t\to\pm\infty}\boldsymbol{\xi}_{j}(t;t_{0},\boldsymbol{x}_{0}),\quad j=1,2,3.\label{eq:Lyapuov exponent-1}
\end{align}
If these limits exist for a given $\left(\boldsymbol{x}_{0},t_{0}\right)$ pair and the forward- and backward-asymptotic behaviors of a trajectory $\boldsymbol{x}(t;t_{0},\boldsymbol{x}_{0})$ are the same (e.g., $\boldsymbol{x}(t;t_{0},\boldsymbol{x}_{0})$ is a periodic orbit, a quasiperiodic torus or a chaotic invariant set), then we have 
\begin{align}
&\mu_{3}^{+}\left(\boldsymbol{x}_{0},t_{0}\right)=\frac{1}{\mu_{1}^{-}\left(\boldsymbol{x}_{0},t_{0}\right)},\quad\mu_{2}^{+}\left(\boldsymbol{x}_{0},t_{0}\right)=\frac{1}{\mu_{2}^{-}\left(\boldsymbol{x}_{0},t_{0}\right)}, \nonumber \\
&\mu_{1}^{+}\left(\boldsymbol{x}_{0},t_{0}\right)=\frac{1}{\mu_{3}^{-}\left(\boldsymbol{x}_{0},t_{0}\right)}, \\
&\boldsymbol{\zeta}_{3}^{+}(\boldsymbol{x}_{0},t_{0})=\boldsymbol{\zeta}_{1}^{-}(\boldsymbol{x}_{0},t_{0}),\quad\boldsymbol{\zeta}_{2}^{+}(\boldsymbol{x}_{0},t_{0})=\boldsymbol{\zeta}_{2}^{-}(\boldsymbol{x}_{0},t_{0}),\nonumber \\
&\boldsymbol{\zeta}_{1}^{+}(\boldsymbol{x}_{0},t_{0})=\boldsymbol{\zeta}_{3}^{-}(\boldsymbol{x}_{0},t_{0}).
\end{align}
In this case, we have
\begin{align}
\boldsymbol{\zeta}_{3}^{+}(\boldsymbol{x}_{0},t_{0})\times\boldsymbol{\zeta}_{3}^{-}(\boldsymbol{x}_{0},t_{0})&=\boldsymbol{\zeta}_{3}^{+}(\boldsymbol{x}_{0},t_{0})\times\boldsymbol{\zeta}_{1}^{+}(\boldsymbol{x}_{0},t_{0}) \nonumber \\
&=\boldsymbol{\zeta}_{3}^{-}(\boldsymbol{x}_{0},t_{0})\times\boldsymbol{\zeta}_{1}^{-}(\boldsymbol{x}_{0},t_{0})
\end{align} As a consequence, if the forward- and backward-asymptotic behaviors along a trajectory are the same, then the vorticity vector $\boldsymbol{\omega}_{t_0}\left(\boldsymbol{x}_{0}\right)$ aligns exactly with the single intermediate Lyapunov vector $\boldsymbol{\zeta}_{2}^{\pm}(\boldsymbol{x}_{0},t_{0})$ in flows satisfying the conditions of Corollary \ref{cor: 1}. 
\end{remark}
\begin{remark}
We now contrast our conclusions with previous vorticity alignment results in the Lagrangian frame. \citet{Ni2014} propose that the evolving vorticity $\boldsymbol{\omega}\left(\boldsymbol{x}(t;t_{0},\boldsymbol{x}_{0}),t\right)$ tends to align  with the dominant left singular vector $\boldsymbol{\eta}_{3}(t;t_{0},\boldsymbol{x}_{0})$ of the deformation gradient $\boldsymbol{\nabla}\mathbf{F}_{t_{0}}^{t}\left(\boldsymbol{x}_{0}\right)$. Their argument is based on the fact that a generic initial condition of the inviscid vorticity transport equation (\ref{eq:inviscid vorticity tranpsort eq}) will align with $\boldsymbol{\nabla}\mathbf{F}_{t_{0}}^{t}\left(\boldsymbol{x}_{0}\right)\boldsymbol{\xi}_{3}(t;t_{0},\boldsymbol{x}_{0})$, given that the initial condition $\boldsymbol{\xi}_{3}(t;t_{0},\boldsymbol{x}_{0})$ stretches the most over the time interval $[t_{0},t]$. \citet{Ni2014} attribute the lack of a perfect alignment between $\boldsymbol{\omega}\left(\boldsymbol{x}(t;t_{0},\boldsymbol{x}_{0}),t\right)$ and $\boldsymbol{\eta}_{3}(t;t_{0},\boldsymbol{x}_{0})$ to viscous effects.  Note, however, that such a perfect alignment cannot take place in inviscid flows (under potential body forces) along trajectories with pointwise uniformly bounded vorticity and positive maximal Lyapunov exponent. The reason is that such an alignment between $\boldsymbol{\omega}\left(\boldsymbol{x}(t;t_{0},\boldsymbol{x}_{0}),t\right)$ and $\boldsymbol{\eta}_{3}(t;t_{0},\boldsymbol{x}_{0})$ would lead to unbounded vorticity growth. Indeed, by Eq.(\ref{eq:identity 3}), the vorticity would grow unbounded in forward time unless $\left\langle \boldsymbol{\omega}_{t_{0}}\left(\boldsymbol{x}_{0}\right),\boldsymbol{\xi}_{3}(t;t_{0},\boldsymbol{x}_{0})\right\rangle =0,$ i.e., unless the vorticity is exactly normal to the dominant stretching direction at all points along a trajectory with positive Lyapunov exponents. This initial alignment then implies that, by the identity (\ref{eq:left-right singular vectors of deformaiton gradient}), the evolving vorticity and the left singular vector must, in fact, satisfy 
\begin{equation}
\boldsymbol{\omega}\left(\boldsymbol{x}(t;t_{0},\boldsymbol{x}_{0}),t\right)\perp\boldsymbol{\eta}_{3}(t;t_{0},\boldsymbol{x}_{0})\label{eq:omega and eta2 orthogonal}
\end{equation}
in an inviscid flow with bounded vorticity, as opposed to $\boldsymbol{\omega}\left(\boldsymbol{x}(t;t_{0},\boldsymbol{x}_{0}),t\right)\parallel\boldsymbol{\eta}_{3}(t;t_{0},\boldsymbol{x}_{0})$.  \\ \\
As most trajectories in a turbulent flow have positive Lyapunov exponents, we conclude for near-inviscid, turbulent flows under potential external forces the following: The alignment of the vorticity over asymptotically long time intervals should be the weakest with the dominant left singular eigenvector $\boldsymbol{\eta}_{3}(t;t_{0},\boldsymbol{x}_{0})$ in comparison with the other two singular vectors. Numerical simulations, such as those of \citet{Ni2014}, may well show substantial alignment with $\boldsymbol{\eta}_{3}(t;t_{0},\boldsymbol{x}_{0})$, but this can
only be the result of numerical instability. Indeed, the slightest
inaccuracy in computing $\boldsymbol{\omega}_{t_{0}}\left(\boldsymbol{x}_{0}\right)$ and the inevitably growing inaccuracies in computing $\boldsymbol{\eta}_{3}(t;t_{0},\boldsymbol{x}_{0})$ could result in a non-zero vorticity component along $\boldsymbol{\eta}_{3}(t;t_{0},\boldsymbol{x}_{0})$, signaling an alignment with $\boldsymbol{\eta}_{3}(t;t_{0},\boldsymbol{x}_{0})$. Such an observed alignment, however, may only be real under more substantial viscous forces for trajectories that in fact enhance, rather than impede, the alignment. 
\end{remark}

\section{Vorticity alignment for Navier-Stokes flows in the Eulerian frame}
\label{sec:EulerianResults}

We first recall the identity 
\[
\left[\boldsymbol{\nabla}\mathbf{v}\right]\boldsymbol{\omega}=\left(\mathbf{S}+\mathbf{W}\right)\boldsymbol{\omega}=\mathbf{S\boldsymbol{\omega}}+\frac{1}{2}\boldsymbol{\omega}\times\mathbf{\boldsymbol{\omega}}=\mathbf{S\boldsymbol{\omega},}
\]
where $\mathbf{W}(\boldsymbol{x},t)=\frac{1}{2}\left[\left[\boldsymbol{\nabla}\mathbf{v}\left(\boldsymbol{x},t\right)\right]^{\mathrm{T}}-\boldsymbol{\nabla}\mathbf{v}\left(\boldsymbol{x},t\right)\right]$ is the spin tensor. Based on this identity, we can re-write the homogeneous linear system of ODEs (\ref{eq:inviscid vorticity tranpsort eq}) as
\begin{equation}
\frac{D\boldsymbol{\omega}}{Dt}=\left[\mathbf{S}\left(\boldsymbol{x}(t;t_{0},\boldsymbol{x}_{0}),t\right)\right]\boldsymbol{\omega}.\label{eq:inviscid vorticity tranpsort eq-1}
\end{equation}
We assume now that along a trajectory, $\boldsymbol{x}(t;t_{0},\boldsymbol{x}_{0})$, the rate-of-strain tensor has an asymptotic mean with uniformly bounded
variation i.e., there exists a function $\boldsymbol{\Gamma}\left(\boldsymbol{x}_{0},t\right)$
such that
\begin{align}
&\mathbf{S}\left(\boldsymbol{x}(t;t_{0},\boldsymbol{x}_{0}),t\right)=\mathbf{S}_{0}(\boldsymbol{x}_{0},t_{0})+\boldsymbol{\Gamma}\left(\boldsymbol{x}_{0},t\right), \notag \\
&\left|\boldsymbol{\Gamma}\left(\boldsymbol{x}_{0},t\right)\right|\leq\gamma\left(\boldsymbol{x}_{0}\right)\ll\infty,\quad t\in\mathbb{R}.\label{eq:rate of strain bounded variation}
\end{align}
The eigenvalues $\sigma_{0j}(\boldsymbol{x}_{0},t_{0})$ and corresponding eigenvectors $ \mathbf{e}_{0j}(\boldsymbol{x}_{0},t_0) $ of $\mathbf{S}_{0}(\boldsymbol{x}_{0},t_{0})$
then satisfy
\begin{equation}
\mathbf{S}_0(\boldsymbol{x}_0,t_0)\mathbf{e}_{0j}(\boldsymbol{x}_{0},t_0) = \sigma_{0j}(\boldsymbol{x}_{0},t_{0})\mathbf{e}_{0j}(\boldsymbol{x}_{0},t_0),\label{eq:e_0j}
\end{equation} 
\begin{equation}
\sigma_{01}(\boldsymbol{x}_{0},t_{0})\leq\sigma_{02}(\boldsymbol{x}_{0},t_{0})\leq\sigma_{03}(\boldsymbol{x}_{0},t_{0}),\qquad\sum_{j=1}^{3}\sigma_{0j}(\boldsymbol{x}_{0},t_{0})=0\label{eq:sigma_0j}.
\end{equation}
Along such a trajectory, the vorticity transport equation (\ref{eq:vorticity transport})
can be re-written as 
\begin{equation}
\frac{D\boldsymbol{\omega}}{Dt}=\mathbf{S}_{0}(\boldsymbol{x}_{0},t_{0})\boldsymbol{\omega}+\nu\Delta\mathbf{\boldsymbol{\omega}}+\boldsymbol{\nabla}\times\mathbf{f}+\boldsymbol{\Gamma}\left(\boldsymbol{x}_{0},t\right)\boldsymbol{\omega},\label{eq:vorticity transport-1}
\end{equation} whose integral form becomes
\begin{align}
\boldsymbol{\omega}_{t}\left(\boldsymbol{x}_{0}\right)&=e^{\mathbf{S}_{0}(\boldsymbol{x}_{0},t_{0})\left(t-t_{0}\right)}\boldsymbol{\omega}_{t_0}\left(\boldsymbol{x}_{0}\right) \\
&+\int_{t_{0}}^{t}e^{\mathbf{S}_{0}(\boldsymbol{x}_{0},t_{0})\left(t-s\right)}\left[ \nu\Delta\boldsymbol{\omega}_s(\boldsymbol{x}_{0}) \right. \nonumber\\
&+ \left.\boldsymbol{\nabla}\times\boldsymbol{f}(\boldsymbol{x}(s;t_0,\boldsymbol{x}_{0}),s)+\boldsymbol{\Gamma}\left(\boldsymbol{x}_0,s\right)\boldsymbol{\omega}_s(\boldsymbol{x}_{0})\right]\,ds.\label{eq:omega(t)-1}
\end{align}
Using this equation in a derivation outlined in the \ref{app: Proof of Theorem 2}, we obtain the following forward-and backward-asymptotic estimates for the alignment between the vorticity and the principal Eulerian stretching directions:
\begin{align}
&\left\langle\mathbf{e}_{\boldsymbol{\omega}_{0}}, \mathbf{e}_{03}(\boldsymbol{x}_{0},t_{0})\right\rangle ^{2} \leq 1- \nonumber \\
&\left(1 - \left|\displaystyle\int_{t_0}^{\infty}e^{\sigma_{03}(\boldsymbol{x}_0,t_0)(t_0-s)} \left\langle \mathbf{e}_{03}(\boldsymbol{x}_0,t_0), \boldsymbol{h}_s(\boldsymbol{x}_0;\nu, \boldsymbol{f})\right\rangle ds  \right|\right)^2 \label{eq:alignment_e3}, \\
&\left\langle\mathbf{e}_{\boldsymbol{\omega}_{0}}, \mathbf{e}_{01}(\boldsymbol{x}_{0},t_{0})\right\rangle ^{2} \leq 1-\nonumber \\ &\left(1 -\left|\displaystyle\int_{t_0}^{-\infty}e^{\sigma_{01}(\boldsymbol{x}_0,t_0)(t_0-s)} \left\langle \mathbf{e}_{01}(\boldsymbol{x}_0,t_0), \boldsymbol{h}_s(\boldsymbol{x}_0;\nu, \boldsymbol{f})\right\rangle ds  \right|\right)^2 \label{eq:alignment_e1},
\end{align} with
\[ \boldsymbol{h}_t(\boldsymbol{x}_0;\nu, \boldsymbol{f}):=\frac{\nu\Delta\boldsymbol{\omega}_t(\boldsymbol{x}_{0})+\boldsymbol{\nabla}\times\boldsymbol{f}(\boldsymbol{x}(t;t_0,\boldsymbol{x}_{0}),t)+\boldsymbol{\Gamma}\left(\boldsymbol{x}_0,t\right)\boldsymbol{\omega}_t(\boldsymbol{x}_{0})}{\left| \boldsymbol{\omega}_{t_0}(\boldsymbol{x}_0)\right|}. \]
The eigenvectors $ \mathbf{e}_{0j}(\boldsymbol{x}_0,t_0) $ are mutually orthogonal and hence
\begin{equation}
\left\langle \mathbf{e}_{01}(\boldsymbol{x}_0,t_0),  \mathbf{e}_{\omega_{t_0}}\right\rangle^2+\left\langle \mathbf{e}_{02}(\boldsymbol{x}_0,t_0),  \mathbf{e}_{\omega_{t_0}}\right\rangle^2+\left\langle \mathbf{e}_{03}(\boldsymbol{x}_0,t_0),  \mathbf{e}_{\omega_{t_0}}\right\rangle^2 = 1.
\end{equation} The angle $ \beta(\boldsymbol{x}_0,t_0) $ between the vorticity $ \boldsymbol{\omega}_0 $ and the intermediate eigenvector $ \mathbf{e}_{02}(\boldsymbol{x}_0,t_0) $ of the rate-of-strain tensor then satisfies
\begin{align}
\sin^2\beta(\boldsymbol{x}_0,t_0)=\left\langle \mathbf{e}_{01}(\boldsymbol{x}_0,t_0),  \mathbf{e}_{\omega_{t_0}}\right\rangle^2+\left\langle \mathbf{e}_{03}(\boldsymbol{x}_0,t_0),  \mathbf{e}_{\omega_{t_0}}\right\rangle^2. \label{eq: sin_beta}
\end{align}
Combining the inequalities (\ref{eq:alignment_e3})-(\ref{eq:alignment_e1}) with Eq.(\ref{eq: sin_beta}) then gives the following strengthened version of Theorem \ref{thm: 1} with respect to the eigenvectors of the rate-of-strain tensor.
\begin{theorem}
\label{thm: 2}Consider an incompressible Navier\textendash Stokes velocity field $\boldsymbol{v}(\boldsymbol{x},t)$, defined on the invariant flow region $U$ with pointwise uniformly bounded vorticity. Assume that along a trajectory $\boldsymbol{x}(t;t_{0},\boldsymbol{x}_{0}),$ the rate-of-strain tensor has a uniformly bounded variation $ \boldsymbol{\Gamma}\left(\boldsymbol{x}_{0},t\right) $ around its initial value $ \mathbf{S}_0(\boldsymbol{x}_0,t_0) $, as in Eq.(\ref{eq:rate of strain bounded variation}). Finally, assume that $ \mathbf{S}_0\left(\boldsymbol{x}_{0},t_{0}\right) $ has no repeated eigenvalues. Then the angle $\beta(\boldsymbol{x}_{0},t_{0})$ between the vorticity $\boldsymbol{\omega}\left(\boldsymbol{x}_{0},t_{0}\right)$ and the intermediate rate-of-strain eigenvector $\mathbf{e}_{02}(\boldsymbol{x}_{0},t_{0})$ satisfies 
\begin{align}
&\sin^2\beta(\boldsymbol{x}_0,t_0) \leq \nonumber \\ & 2 -\left(1-\left|\displaystyle\int_{t_0}^{\infty}e^{\sigma_{03}(\boldsymbol{x}_0,t_0)(t_0-s)} \left\langle \mathbf{e}_{03}(\boldsymbol{x}_0,t_0), \boldsymbol{h}_s(\boldsymbol{x}_0;\nu, \boldsymbol{f})\right\rangle ds  \right|\right)^2 \nonumber \\
&-\left(1 -\left|\displaystyle\int_{t_0}^{-\infty}e^{\sigma_{01}(\boldsymbol{x}_0,t_0)(t_0-s)} \left\langle \mathbf{e}_{01}(\boldsymbol{x}_0,t_0), \boldsymbol{h}_s(\boldsymbol{x}_0;\nu, \boldsymbol{f})\right\rangle ds  \right|\right)^2\label{eq:near-alignment-1}.
\end{align}
\end{theorem}
By writing out the right-hand side of inequality (\ref{eq:near-alignment-1}), we can further upper estimate $ \sin^2\beta(\boldsymbol{x}_0,t_0) $ as
\begin{align}
&\sin^2\beta(\boldsymbol{x}_0,t_0) \leq \nonumber \\ & 2\left(\left|\displaystyle\int_{t_0}^{\infty}e^{\sigma_{03}(\boldsymbol{x}_0,t_0)(t_0-s)} \left\langle \mathbf{e}_{03}(\boldsymbol{x}_0,t_0), \boldsymbol{h}_s(\boldsymbol{x}_0;\nu, \boldsymbol{f})\right\rangle ds  \right| \right. \nonumber\\
& +\left. \left|\displaystyle\int_{t_0}^{-\infty}e^{\sigma_{01}(\boldsymbol{x}_0,t_0)(t_0-s)} \left\langle \mathbf{e}_{01}(\boldsymbol{x}_0,t_0), \boldsymbol{h}_s(\boldsymbol{x}_0;\nu, \boldsymbol{f})\right\rangle ds  \right|\right)
\end{align}
Therefore, the upper bound from inequality (\ref{eq:near-alignment-1}) becomes tight if the forward and backward integrals are sufficiently small
\begin{align}  \left|\displaystyle\int_{t_0}^{\infty}e^{\sigma_{03}(\boldsymbol{x}_0,t_0)(t_0-s)} \left\langle \mathbf{e}_{03}(\boldsymbol{x}_0,t_0), \boldsymbol{h}_s(\boldsymbol{x}_0;\nu, \boldsymbol{f})\right\rangle ds  \right| &\ll 1, \\ \left|\displaystyle\int_{t_0}^{-\infty}e^{\sigma_{01}(\boldsymbol{x}_0,t_0)(t_0-s)} \left\langle \mathbf{e}_{01}(\boldsymbol{x}_0,t_0), \boldsymbol{h}_s(\boldsymbol{x}_0;\nu, \boldsymbol{f})\right\rangle ds  \right| &\ll 1.
\end{align}
Physically,  the integrals from Theorem \ref{thm: 2} represent a weighted averaged, i.e. convolution, between the exponential of the principal strain rates $ \left(\sigma_{03}(\boldsymbol{x}_0)>0,\sigma_{01}(\boldsymbol{x}_0)<0\right) $ and the sum of the strain rate variation times the vorticity and the curl of the viscous and non-conservative forces normalized by the initial vorticity magnitude.  The exponential functions respectively vanish in the asymptotic limits $ t \rightarrow \pm \infty  $ and the exponential decay is determined by the exponents $ \sigma_{03}(\boldsymbol{x}_0) $ and $ \sigma_{01}(\boldsymbol{x}_0) $.  The asymptotic estimates in equations (\ref{eq:alignment_e3})-(\ref{eq:alignment_e1}) are therefore monotonically decreasing functions of $ \sigma_{03}(\boldsymbol{x}_0) $ and $ -\sigma_{01}(\boldsymbol{x}_0) $.  High stretching and compression in the $ \mathbf{e}_{03}(\boldsymbol{x}_0)-\mathbf{e}_{01}(\boldsymbol{x}_0) $ results in a quick exponential decay of the upper bounds (\ref{eq:alignment_e3})-(\ref{eq:alignment_e1}).  Under the conditions of Theorem \ref{thm: 2}, we have therefore obtained that the vorticity vector approximately aligns with the intermediate rate-of-strain eigenvector in high strain regions provided that the curl of the viscous and external forces and the strain rate variation along a short trajectory segment are small relative to the initial vorticity magnitude. We expect these conditions to hold for a large number of trajectories in isotropic turbulent flows \citep{Girimaji1990}.  \\ \\ 
For inviscid flows under potential forces, we then obtain the following strengthened version of Theorem \ref{thm: 2}.
\begin{corollary}
\label{cor: 2}
In inviscid flows defined on a compact invariant domain $ U $ with pointwise uniformly bounded vorticity and potential forces,  the angle $ \beta(\boldsymbol{x}_0,t_0) $ between  the vorticity $ \boldsymbol{\omega}_0 $ and the intermediate eigenvector $ \mathbf{e}_{02}(\boldsymbol{x}_0,t_0) $ of the initial rate-of-strain tensor $ \mathbf{S}_0(\boldsymbol{x}_0,t_0) $ satisfies
\begin{align}
&\sin^2\beta(\boldsymbol{x}_0,t_0) \leq \nonumber \\
&2-\left(1-\left|\displaystyle\int_{t_0}^{\infty}e^{\sigma_{03}(\boldsymbol{x}_0,t_0)(t_0-s)} \left\langle \mathbf{e}_{03}(\boldsymbol{x}_0,t_0), \mathbf{\Gamma}(\boldsymbol{x}_0,s)\frac{\omega_s(\boldsymbol{x}_0)}{\left|\omega_{t_0}(\boldsymbol{x}_0)\right|}\right\rangle ds  \right|\right)^2 \nonumber \\
&-\left(1 -\left|\displaystyle\int_{t_0}^{-\infty}e^{\sigma_{01}(\boldsymbol{x}_0,t_0)(t_0-s)} \left\langle \mathbf{e}_{01}(\boldsymbol{x}_0,t_0), \mathbf{\Gamma}(\boldsymbol{x}_0,s)\frac{\omega_s(\boldsymbol{x}_0)}{\left|\omega_{t_0}(\boldsymbol{x}_0)\right|}\right\rangle ds  \right|\right)^2 \label{eq:AlignmentEulerian-corollary3}.
\end{align}
\end{corollary}
In steady flows, the rate-of-strain tensor has no explicit time-dependence and the variation of the rate-of-strain tensor is identically zero at fixed points. For such flows, therefore, Corollary \ref{cor: 2} implies that at fixed points, the vorticity aligns exactly with the intermediate eigenvector of the rate-of-strain tensor $ \mathbf{S}_0(\boldsymbol{x}_0) $ since the right-hand side vanishes at such points. At fixed points we thus have $ \sin^2{\beta}(\boldsymbol{x}_0,t_0) \equiv 0 $.  Additionally, $ \sin^2\beta(\boldsymbol{x}_0,t_0) $ depends continuously on the initial conditions $ \boldsymbol{x}_0 $. Therefore, the preferential alignment between $ \boldsymbol{\omega}_{t_0} $ and $ \mathbf{e}_{02}(\boldsymbol{x}_0) $ extends to a sufficiently small neighbourhood around the fixed point.
\section{Numerical Examples}
We first investigate the alignment properties of the vorticity in an analytic non-integrable 3D Euler flow with chaotic trajectories.  We then proceed to analyze vorticity alignment in 3D homogenous isotropic turbulence data from the Johns Hopkins Turbulence Database \citep{Li2008}.
\subsection{Steady ABC-flow}
\begin{figure*}
\centering \includegraphics[width=1\textwidth]{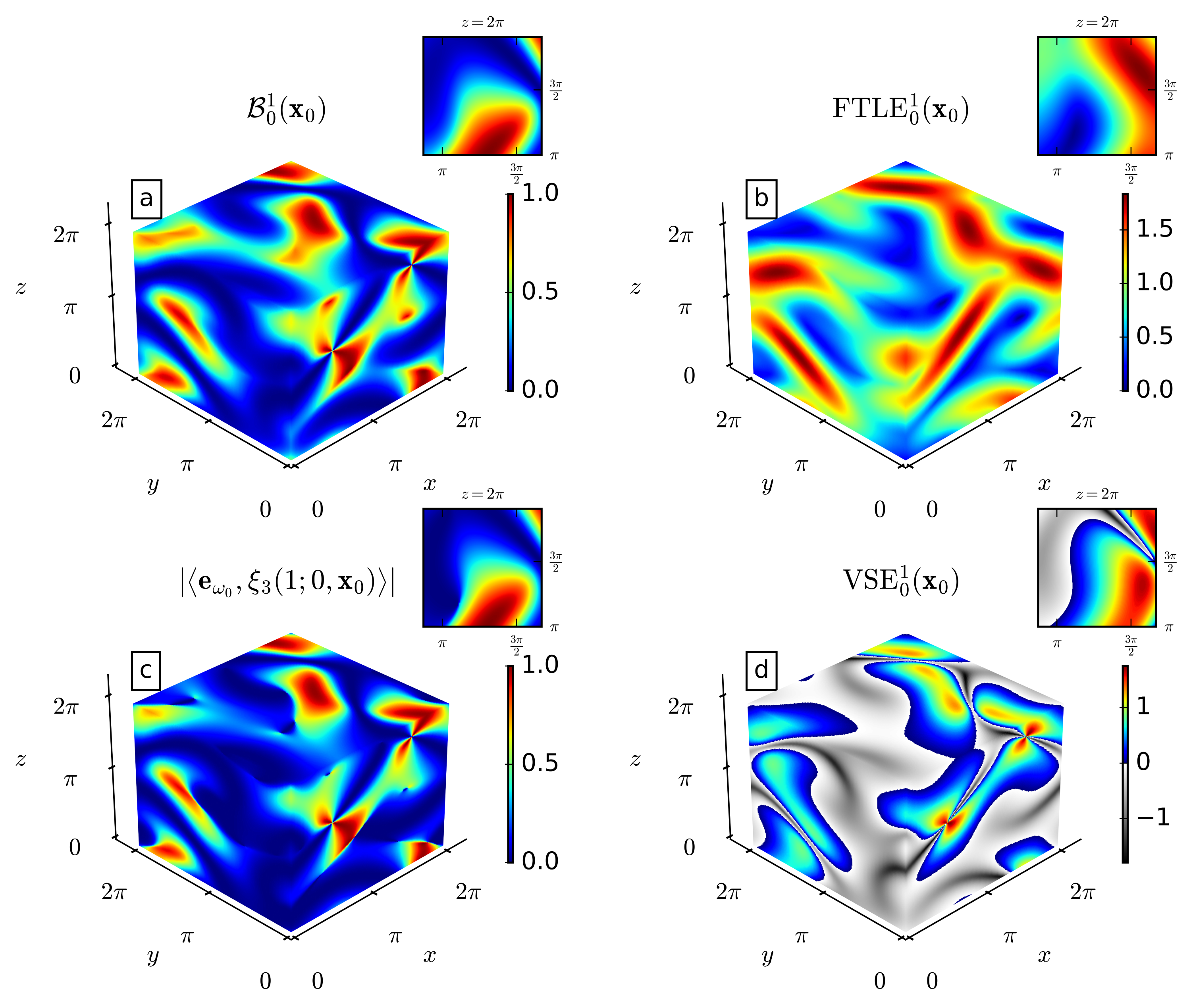} \caption{Vorticity alignment and Lagrangian diagnostics for the ABC-flow for $ t \in [0,1] $. 
\label{fig:Fig1}}
\end{figure*}

\begin{figure*}
\centering \includegraphics[width=1\textwidth]{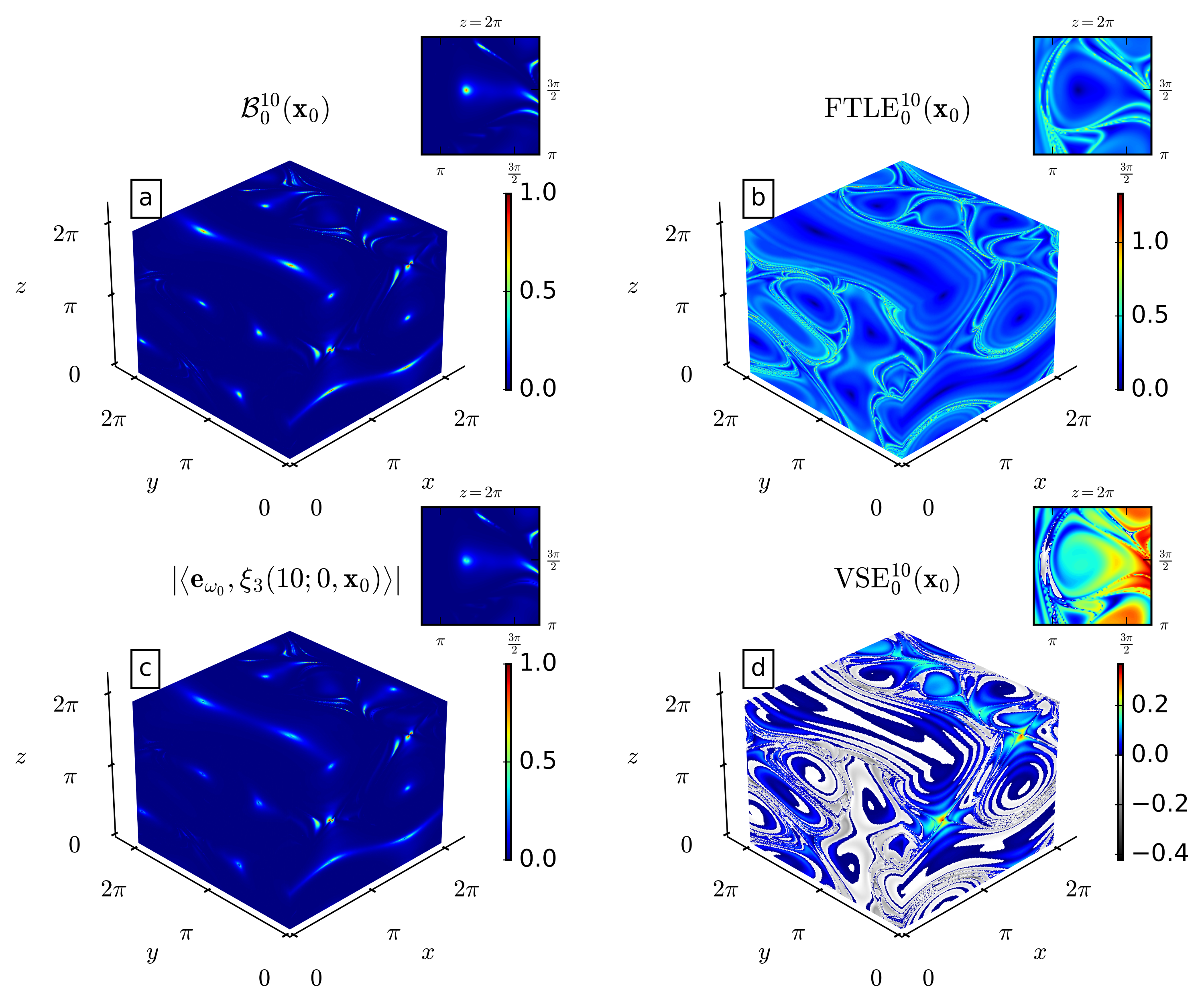} \caption{Same as Fig. \ref{fig:Fig1} but over the time interval $ [0,10] $.
\label{fig:Fig2}}
\end{figure*}

\begin{figure*}
\centering
\captionsetup{width=1\linewidth}
\includegraphics[width=1\textwidth]{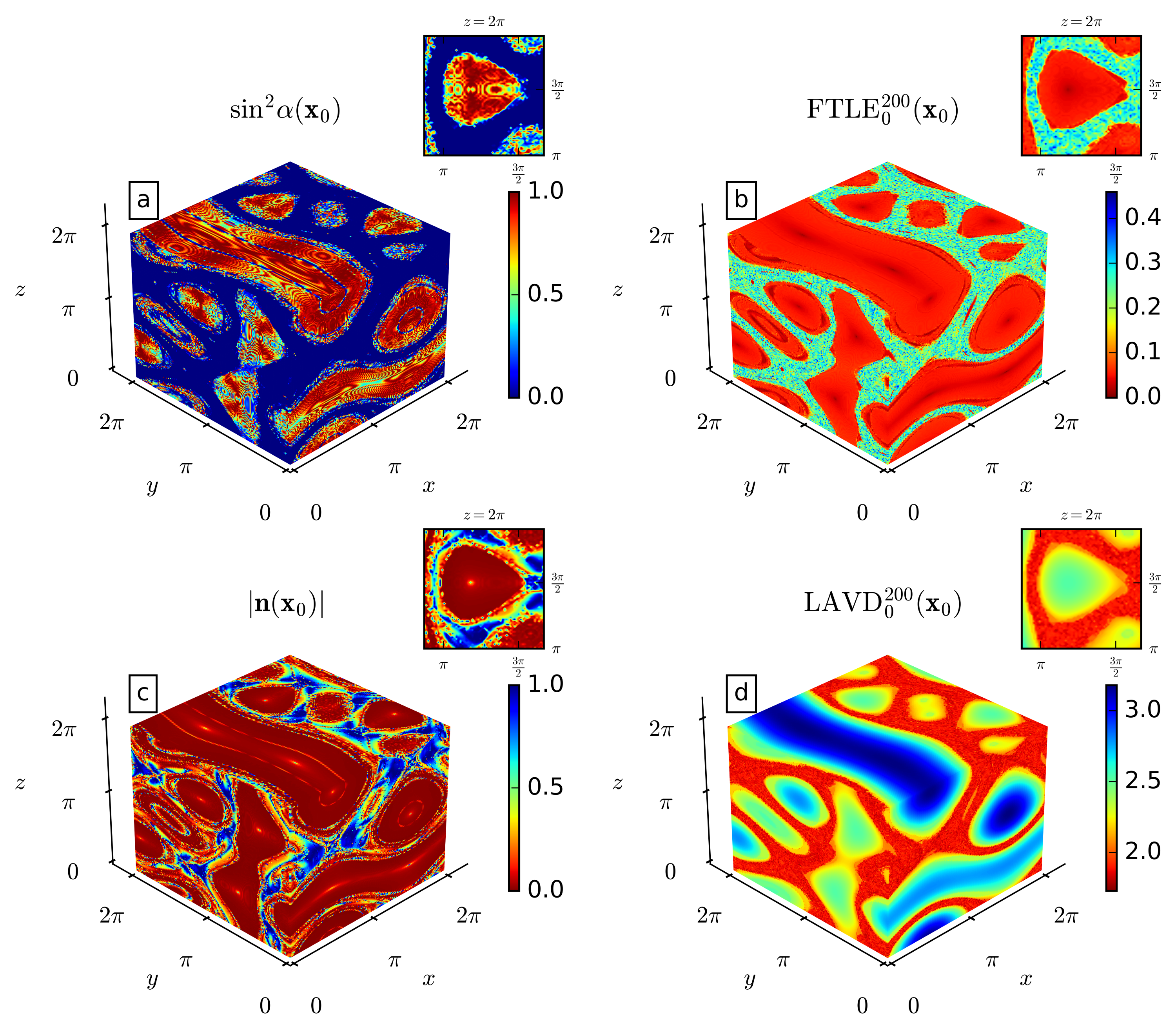} \caption{Alignment diagnostics vs.  hyperbolic and elliptic regions in the ABC flow. The zoomed insets show an elliptic region with invariant tori surrounded by hyperbolic region with chaotic dynamics.
\label{fig:Fig3}}
\end{figure*}

\begin{figure*}
\centering \includegraphics[width=1\textwidth]{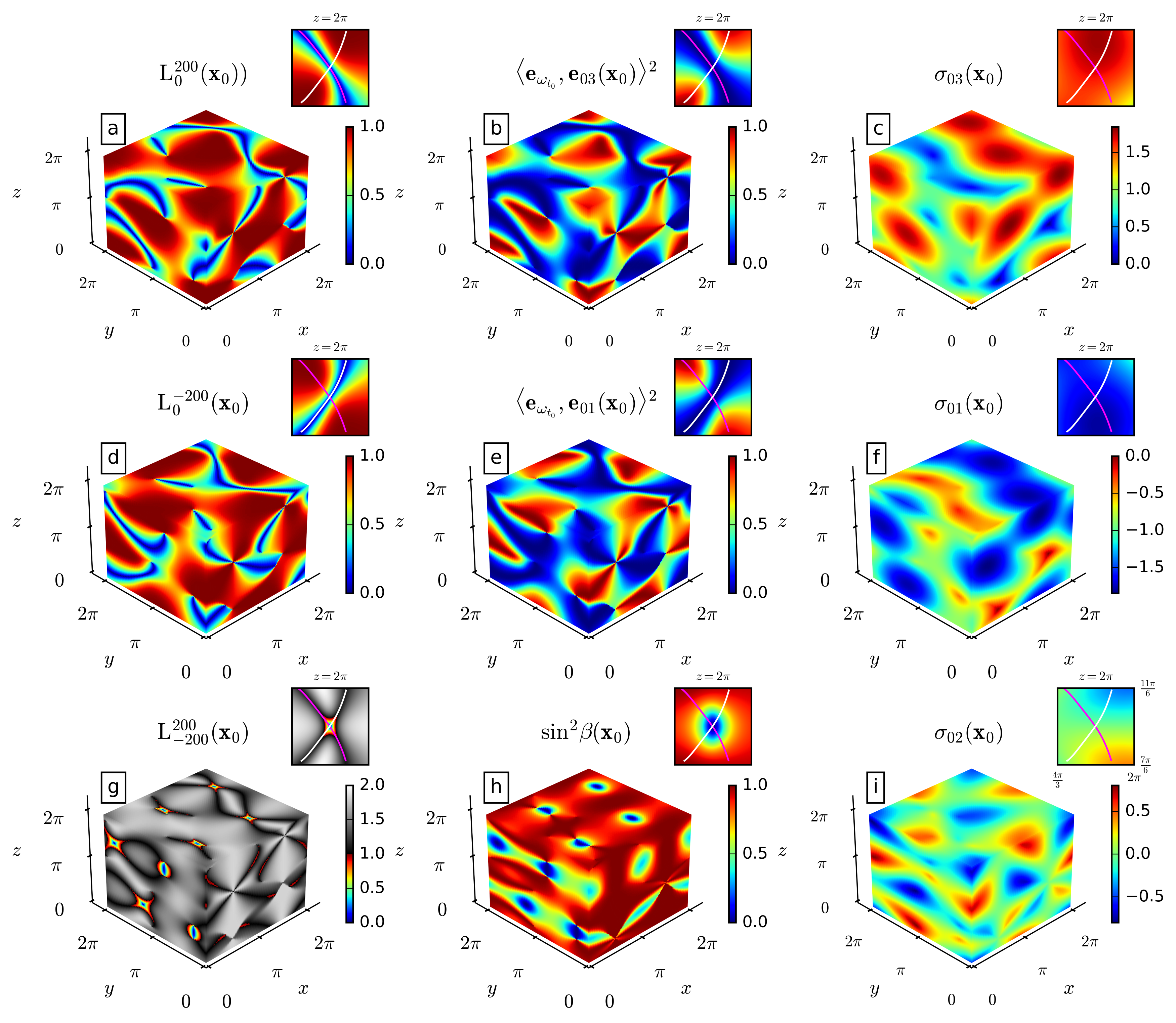} \caption{Vorticity alignment and Eulerian diagnostics for the ABC flow.
\label{fig:Fig4}}
\end{figure*}

As a first example, we consider the fully 3D  steady Arnold-Beltrami-Childress (ABC) flow \citep{Dombre1986}, given by the velocity field
\begin{equation}
\boldsymbol{v}(\boldsymbol{x}) = \begin{pmatrix}
C\sin(z)+ B\cos(y) \\ A \sin(x)+ C\cos(z) \\ B \sin(y)+A\cos(x)
\end{pmatrix},
\end{equation} defined on the 3D toroidal domain $ U \in [0, 2\pi]^3 $ with $ A=\sqrt{2} ,  B = 1, C=\sqrt{3} $. This velocity field is an exact solution of the incompressible Euler equation. The flow shows a very rich behaviour and admits a variety of features analogous to KAM tori and hyperbolic structures \citep{Haller2001,Blazevski2014}. The ABC-flow is a Beltrami flow that satisfies:
\begin{equation}
\boldsymbol{v}(\boldsymbol{x}) = \boldsymbol{\omega}(\boldsymbol{x}),
\end{equation} and hence the vorticity is globally bounded.
Since $ \boldsymbol{v}(\boldsymbol{x}) $ is a steady velocity field defined on a compact domain given by the three-dimensional torus, Oseledets multiplicative ergodic theorem guarantees the existence of the Lyapunov exponents $ \mu_3^{\pm}(\boldsymbol{x}_0) $ and the Lyapunov vectors $ \boldsymbol{\zeta}_3^{\pm}(\boldsymbol{x}_0) $ for almost all initial conditions $ \boldsymbol{x}_0 $. The finite-time Lyapunov exponent and the right singular vectors $ \boldsymbol{\xi}_3 (\boldsymbol{x}_0)$ of the linearized flow map are thus guaranteed to converge for nearly all initial conditions  to $ \mu_3^{\pm}(\boldsymbol{x}_0) $ and $ \boldsymbol{\zeta}_3^{\pm}(\boldsymbol{x}_0) $. We will compute the linearized flow map $ \nabla \mathbf{F}_{t_0}^{t_1}(\boldsymbol{x}_0) $ along a trajectory launched from $ \boldsymbol{x}_0 $ over a sufficiently long time interval $ t \in [t_0,t_1] $ and then obtain $ \mathrm{FTLE}_{t_0}^{t_1}(\boldsymbol{x}_0) $ and $ \boldsymbol{\xi}_3(t_1;t_0, \boldsymbol{x}_0) $ from a singular value decomposition (SVD) of $ \nabla \mathbf{F}_{t_0}^{t_1}(\boldsymbol{x}_0) $, as done by \citet{Oettinger2016}. \\ \\
Figures (\ref{fig:Fig1}a-\ref{fig:Fig2}a) respectively show the $ \mathcal{B}_{t_0}^{t_1} $ field (defined in 
Eq.(\ref{eq: cal B})), computed over the time intervals $ [0,1] $ and $ [0,10] $. Features of the $ \mathcal{B}_{t_0}^{t_1} $ field are related to those of the $ \mathrm{FTLE}_{t_0}^{t_1} $ field for sufficiently small $ \mathrm{VSE}_{t_0}^{t_1} $, as seen from the formula (\ref{eq: cal B}). Negative trenches of the  $ \mathrm{VSE}_{t_0}^{t_1} $ field indicate areas where the vorticity vector is compressed, whereas positive ridges of the $ \mathrm{VSE}_{t_0}^{t_1} $ field indicate vortex stretching. Minima in the $ \mathcal{B}_{t_0}^{t_1} $ field reveal areas where the vorticity is preferentially orthogonal to the most stretching direction given by $ \mathbf{\xi}_3(t_1;t_0,\boldsymbol{x}_0) $. The zoomed insets in Fig. \ref{fig:Fig1} show that for short intervals, vortex stretching ($ \mathrm{VSE} > 0 $) is caused by the initial alignment of $ \boldsymbol{\omega}_0 $ with the most stretching direction $ \boldsymbol{\xi}_3 $. On the contrary compression of the vorticity vector ($ \mathrm{VSE} < 0 $) is associated with orthogonality between $ \boldsymbol{\omega}_0 $ and $ \boldsymbol{\xi}_3 $. The contribution of $ \mathrm{VSE}_{t_0}^{t_1} $ to the $ \mathcal{B}_{t_0}^{t_1} $ field becomes negligible as time increases because the vorticity remains bounded (see also zoomed insets of Fig. \ref{fig:Fig2}). \\ \\
Taking the asymptotic limits $ t_1 \rightarrow \pm \infty $, we therefore conclude the following from Corollary \ref{cor: 1}: For almost all initial conditions with positive Lyapunov exponents, the vorticity exactly aligns with \[ \boldsymbol{n}(\boldsymbol{x}_0) = \boldsymbol{\zeta}_3^{+}(\boldsymbol{x}_0) \times \boldsymbol{\zeta}_3^{-}(\boldsymbol{x}_0), \] i.e., with the direction along the intersection of the two planes orthogonal to the forward and backward dominant Lyapunov vectors $ \boldsymbol{\zeta}_3^{\pm}(\boldsymbol{x}_0) $. \\ \\
If the forward and backward most stretching directions are collinear ($ \boldsymbol{\zeta}_3^{+}(\boldsymbol{x}_0) \parallel  \boldsymbol{\zeta}_3^{-}(\boldsymbol{x}_0) $), then there is no unique intersection line between the two planes. In that case the angle  $ \alpha(\boldsymbol{x}_0) $ between $ \boldsymbol{\omega}_0 $ and $ \boldsymbol{n}(\boldsymbol{x}_0) $ is not well defined. Away from such degenerate points, $ \sin^2\alpha(\boldsymbol{x}_0) $ is given by
\begin{align}
\sin^2\alpha(\boldsymbol{x}_0) &= \frac{\left| \mathbf{e}_{\boldsymbol{\omega}_{t_0}} \times \boldsymbol{n}(\boldsymbol{x}_0)\right|^2}{\left| \boldsymbol{n}(\boldsymbol{x}_0) \right|^2} \nonumber \\
&= \frac{\left| \mathbf{e}_{\boldsymbol{\omega}_{t_0}} \times \left(\boldsymbol{\zeta}_3^{+}(\boldsymbol{x}_0) \times \boldsymbol{\zeta}_3^{-}(\boldsymbol{x}_0)\right) \right|^2}{\left| \boldsymbol{\zeta}_3^{+}(\boldsymbol{x}_0) \times \boldsymbol{\zeta}_3^{-}(\boldsymbol{x}_0) \right|^2} \nonumber \\
&= \frac{\left(\left\langle \mathbf{e}_{\boldsymbol{\omega}_{t_0}}, \boldsymbol{\zeta}_3^{+}(\boldsymbol{x}_0) \right\rangle \boldsymbol{\zeta}_3^{-}(\boldsymbol{x}_0)-\left\langle \mathbf{e}_{\boldsymbol{\omega}_{t_0}}, \boldsymbol{\zeta}_3^{-}(\boldsymbol{x}_0) \right\rangle \boldsymbol{\zeta}_3^{+}(\boldsymbol{x}_0) \right)^2}{\left| \boldsymbol{\zeta}_3^{+}(\boldsymbol{x}_0) \times \boldsymbol{\zeta}_3^{-}(\boldsymbol{x}_0) \right|^2} \nonumber \\
&= \frac{\left\langle \mathbf{e}_{\boldsymbol{\omega}_{t_0}}, \boldsymbol{\zeta}_3^{+}(\boldsymbol{x}_0) \right\rangle^2+\left\langle \mathbf{e}_{\boldsymbol{\omega}_{t_0}}, \boldsymbol{\zeta}_3^{-}(\boldsymbol{x}_0) \right\rangle^2}{\left| \boldsymbol{n}(\boldsymbol{x}_0) \right|^2}
\end{align}
This formula shows $ \sin^2\alpha(\boldsymbol{x}_0) $ to be inversely proportional to $ \left| \boldsymbol{n}(\boldsymbol{x}_0)\right|^2 $. Therefore, in regions where the forward and backward dominant Lyapunov vectors are nearly collinear $\left( \left| \boldsymbol{n}(\boldsymbol{x}_0)\right|^2 \ll 1 \right) $, the numerical computation of $ \sin^2\alpha(\boldsymbol{x}_0) $ will be noisy.  \\ \\
Figure \ref{fig:Fig3} displays $ \sin^2\alpha(\boldsymbol{x}_0) $ together with the FTLE and the Lagrangian Averaged Vorticity Deviation (LAVD) given by:
\begin{equation}
\mathrm{LAVD}_{t_0}^t(\boldsymbol{x}_0) = \dfrac{1}{\left|t-t_0\right|} \displaystyle\int_{t_0}^t\left|\boldsymbol{\omega}_s(\boldsymbol{x}_0)-\overline{\boldsymbol{\omega}_s(\boldsymbol{x}_0)}\right|ds,
\end{equation} where $ \overline{\mathbf{\omega}_s(\boldsymbol{x}_0)} $ denotes the spatially averaged vorticity.  The LAVD is an objective diagnostic used to visualize vortical flow structures from three-dimensional velocity data \citep{Haller2016}.  We approximate the leading Lyapunov exponent over a finite time interval as \[ \mu_3^+(\boldsymbol{x}_0) \sim \mathrm{FTLE}_{0}^{200}(\boldsymbol{x}_0). \]  Elliptic (or vortical) flow structures are shear dominated areas characterized by high LAVD-values.  Material elements stretch algebraically within those regions and we therefore expect a vanishing leading Lyapunov exponent \citep{Haller2012}. In contrast,  hyperbolic flow areas are characterized by positive leading Lyapunov exponents because of exponentially stretching material lines. \\ \\
Figure \ref{fig:Fig3} reveals that the forward and backward leading Lyapunov vectors are geometrically nearly collinear within elliptic flow regions (see Fig. \ref{fig:Fig3}c-d). By approximating $ \mu_3^+(\boldsymbol{x}_0) $ from a finite-time computation, we numerically show that Lyapunov exponents vanish within elliptic flow areas, i.e. generalized KAM-tori. Our computations on the steady ABC flow highlight that the vorticity aligns with $ \boldsymbol{n}(\boldsymbol{x}_0) $ in hyperbolic flow regions characterized by positive leading Lyapunov exponents (see Fig. \ref{fig:Fig3}a-b) as predicted by Theorem \ref{thm: 1}. \\ \\
In Figure \ref{fig:Fig4} we show the alignment between the vorticity and the eigenvectors of the rate\textendash of\textendash strain tensor. We approximate the asymptotic estimates (\ref{eq:alignment_e3})-(\ref{eq:near-alignment-1}) from a forward and backward computation over a time interval $ t \in [0,\pm 200] $ and define 
\begin{align}
&\mathrm{L}_0^{200}(\boldsymbol{x}_0) = 1-\left(1- \right. \nonumber \\
&\left. \left|\displaystyle\int_{0}^{200}e^{-\sigma_{03}(\boldsymbol{x}_0,t_0)s} \left\langle \mathbf{e}_{03}(\boldsymbol{x}_0,t_0), \mathbf{\Gamma}(\boldsymbol{x}_0,s)\frac{\omega_s(\boldsymbol{x}_0)}{\left|\omega_{t_0}(\boldsymbol{x}_0)\right|}\right\rangle ds  \right|\right)^2, \\
&\mathrm{L}_0^{-200}(\boldsymbol{x}_0) = 1-\left(1- \right. \nonumber \\
& \left. \left|\displaystyle\int_{0}^{-200}e^{-\sigma_{01}(\boldsymbol{x}_0,t_0)s} \left\langle \mathbf{e}_{01}(\boldsymbol{x}_0,t_0), \mathbf{\Gamma}(\boldsymbol{x}_0,s)\frac{\omega_s(\boldsymbol{x}_0)}{\left|\omega_{t_0}(\boldsymbol{x}_0)\right|}\right\rangle ds  \right|\right)^2, \\
&\mathrm{L}_{-200}^{200}(\boldsymbol{x}_0)=\mathrm{L}_0^{-200}(\boldsymbol{x}_0)+\mathrm{L}_0^{200}(\boldsymbol{x}_0).
\end{align} According to Corollary \ref{cor: 2}, the vorticity preferentially aligns with the intermediate eigenvector $ \mathbf{e}_{02}(\boldsymbol{x}_0) $ of the rate-of-strain tensor $ \mathbf{S}_0(\boldsymbol{x}_0)=\mathbf{S}(\boldsymbol{x}_0) $ given that $ \mathrm{L}_{-200}^{200}(\boldsymbol{x}_0) $ is sufficiently small. Contrary to common observations for homogenous isotropic turbulence, the vorticity shows preferential alignment with $ \mathbf{e}_{02}(\boldsymbol{x}_0) $ in the ABC flow only for very small regions. This region coincides with areas where $ \mathrm{L}_{-200}^{200}(\boldsymbol{x}_0) $ is minimal (see zoomed insets of Fig. \ref{fig:Fig4}).  The zoomed insets additionally display branches of unstable (white line) and stable (magenta line) manifolds, which are approximated through ridges of the backward and forward FTLE field.  At the intersection between stable and unstable manifolds, the vorticity strongly aligns with the intermediate eigenvectors of the rate-of-strain tensor. \\ \\
Therefore, the Eulerian alignment observed in homogenous isotropic turbulence generally does not carry over to inviscid flows with bounded vorticity. Instead,  the vorticity is exactly orthogonal to the forward and backward leading Lyapunov vectors in unforced inviscid flows along trajectories with positive forward and backward Lyapunov exponents. 
\subsection{Forced Homogenous Isotropic Turbulence from Johns Hopkins Turbulence Database}
\begin{figure*}
\centering \includegraphics[width=1\textwidth]{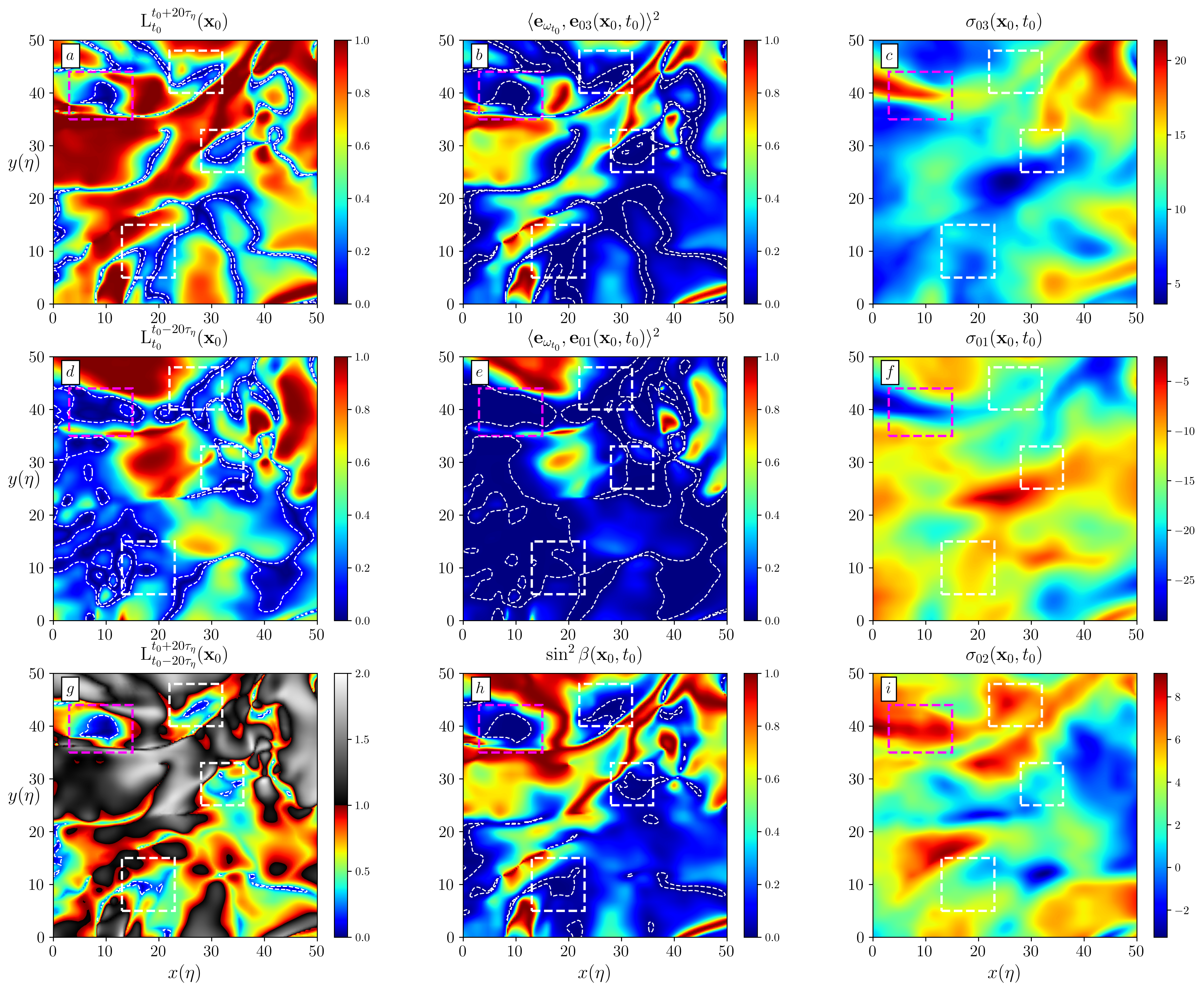} \caption{Alignment in the homogeneous isotropic turbulence data set between the vorticity and the eigenvectors of the rate-of-strain tensor.  
\label{fig:Fig6}}
\end{figure*}

\begin{figure*}
\centering 
\captionsetup{width=1\linewidth}\includegraphics[width=1\textwidth]{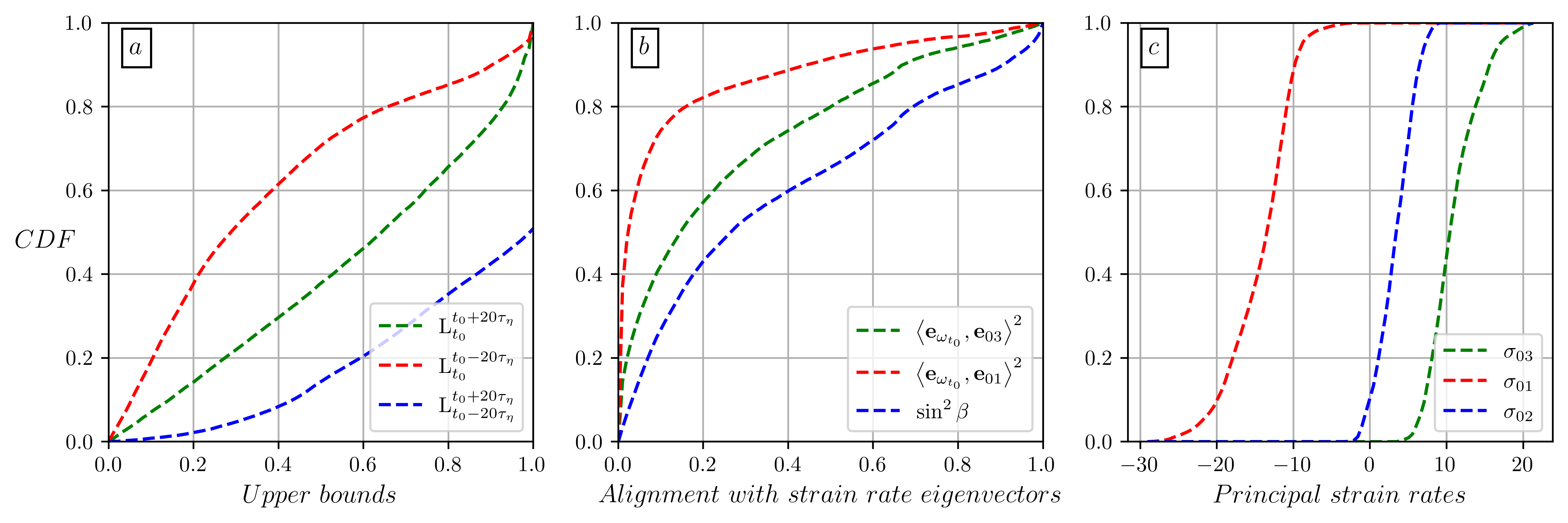} \caption{CDF of the upper bounds defined in equations (\ref{eq: L_+_JHDTB})-(\ref{eq: L_JHDTB}) (panel a). CDF of $ \left\langle \mathbf{e}_{\omega_{t_0}}, \mathbf{e}_{01}(\boldsymbol{x}_0,t_0)\right\rangle^2 $, $ \left\langle \mathbf{e}_{\omega_{t_0}}, \mathbf{e}_{03}(\boldsymbol{x}_0,t_0)\right\rangle^2 $ and $ \sin^2\beta(\boldsymbol{x}_0,t_0)=1-\left\langle \mathbf{e}_{\omega_{t_0}}, \mathbf{e}_{02}(\boldsymbol{x}_0,t_0)\right\rangle^2 $. CDF of the principal strain rates (panel c).
\label{fig:Fig7}}
\end{figure*}

Next, we investigate vorticity alignment in forced homogenous isotropic turbulence from the Johns Hopkins Turbulence Database (JHTDB) \citep{Li2008}. The isotropic turbulence fields were obtained from a $1024^3$-node direct numerical simulation (DNS) that is publicly available from the JHTDB over a periodic box of size $ [0,2\pi]^3 $. The Taylor-Reynolds number fluctuates around $ Re_{\lambda} \sim 433 $. In order to keep the velocity field at a statistically stationary state, the flow is externally forced at large scales by keeping the total energy constant in modes such that their wave number magnitude is less than or equal to 2. The main parameters of the simulation are given in Table \ref{tab: JHTDB}.
\begin{table}[h]
\centering
\begin{tabular}{|c||c|c|c|c|c|}
    \hline
    Dataset & Nodes & $ \nu $ & $ \eta $ & $ \tau_{\eta} $ & $Re_{\lambda}$\\
    \hline  \hline
    JHTDB & $ 1024^3 $ & $ 1.85 \cdot 10^{-4} $ & $ 2.87 \cdot 10^{-3} $ & $ 0.045 $ & $ 433 $ \\
    \hline
\end{tabular}
\caption{Parameters for the data from the homogenous isotropic turbulence simulation of the Johns Hopkins Turbulence Database (JHTDB)}
\label{tab: JHTDB}
\end{table} \\
The Lagrangian trajectories and the velocity gradients are publicly accessible through a web-based interface. The total simulation time of the stored trajectories is 20 Kolmogorov time scales ($ \tau_{\eta} $), with the data provided at a time step of $ 0.1\tau_{\eta} $. Once the trajectories are calculated, the velocity gradients are computed at each point using a 4th-order central finite-difference method with 4th-order Lagrange interpolation.  We select a 2D slice over a domain-size given by $ 50 \eta \times 50 \eta  $, where $ \eta $ is the Kolmogorov length scale. For this study, we use $ 150 \times 150 = 22500 $ Lagrangian particle trajectories to compute the relevant statistical quantities. The viscous and non-conservative forces are computed from the vorticity evolution according to Eq.(\ref{eq:vorticity transport}).
We approximate the asymptotic estimates (\ref{eq:alignment_e3})-(\ref{eq:near-alignment-1}) over a long but finite time interval $ [t_0,t_0\pm20 \tau_{\eta}] $ through
\begin{align}
&\mathrm{L}_{t_0}^{t_0+20\tau_{\eta}}(\boldsymbol{x}_0) = 1-\left(1 - \right. \nonumber\\
& \left. \left|\displaystyle\int_{t_0}^{t_0+20\tau_{\eta}}e^{\sigma_{03}(\boldsymbol{x}_0,t_0)(t_0-s)} \left\langle \mathbf{e}_{03}(\boldsymbol{x}_0,t_0), \boldsymbol{h}_s(\boldsymbol{x}_0;\nu, \boldsymbol{f})\right\rangle ds  \right|\right)^2, \label{eq: L_+_JHDTB}\\
&\mathrm{L}_{t_0}^{t_0-20\tau_{\eta}}(\boldsymbol{x}_0)  = 1-\left(1 - \right. \nonumber \\
& \left. \left|\displaystyle\int_{t_0}^{t_0-20\tau_{\eta}}e^{\sigma_{01}(\boldsymbol{x}_0,t_0)(t_0-s)} \left\langle \mathbf{e}_{01}(\boldsymbol{x}_0,t_0), \boldsymbol{h}_s(\boldsymbol{x}_0;\nu, \boldsymbol{f})\right\rangle ds  \right|\right)^2, \label{eq: L_-_JHDTB}\\
&\mathrm{L}_{t_0-20\tau_{\eta}}^{t_0+20\tau_{\eta}}(\boldsymbol{x}_0)  = \mathrm{L}_{t_0}^{t_0+20\tau_{\eta}}(\boldsymbol{x}_0) +\mathrm{L}_{t_0}^{t_0-20\tau_{\eta}}(\boldsymbol{x}_0)\label{eq: L_JHDTB}.
\end{align}
In Figure \ref{fig:Fig6} we plot the upper bounds defined in equations (\ref{eq: L_+_JHDTB})-(\ref{eq: L_JHDTB}) (panels a,d,g), the alignments between $ \boldsymbol{\omega}_{t_0} $ and the eigenvectors of the rate-of-strain tensor $ \mathbf{S}_{0}(\boldsymbol{x}_0,t_0) $  (panels b,e,h), and the principal strain rates $ \sigma_{0j}(\boldsymbol{x}_0,t_0) $ (panels c,f,i) over the initial conditions $ \boldsymbol{x} _0$. In panel g, we have used a different colormap for $ \mathrm{L}_{t_0-20\tau_{\eta}}^{t_0+20\tau_{\eta}}(\boldsymbol{x}_0) > 1 $. Trenches of the $ \mathrm{L}_{t_0}^{t_0+20\tau_{\eta}}(\boldsymbol{x}_0) $ and $ \mathrm{L}_{t_0}^{t_0-20\tau_{\eta}}(\boldsymbol{x}_0) $ fields indicate areas where the vorticity is preferentially orthogonal to $ \mathbf{e}_{03}(\boldsymbol{x}_0,t_0) $ or $ \mathbf{e}_{01}(\boldsymbol{x}_0,t_0) $. By combining the forward and the backward estimates, we obtain that low values of the $ \mathrm{L}_{t_0-20\tau}^{t_0+20\tau}(\boldsymbol{x}_0) $ field guarantee strong alignment between $ \boldsymbol{\omega}_{t_0} $ and $ \mathbf{e}_{02}(\boldsymbol{x}_0) $.  Furthermore, the dashed white contours emphasize that the upper bounds (panels a,d,g) are topologically similar to the corresponding alignment fields (panels b,e,h). \\ \\ 
The magenta inset in figure \ref{fig:Fig6} highlights a region where the estimate $ \mathrm{L}_{t_0-20\tau_{\eta}}^{t_0+20\tau_{\eta}}(\boldsymbol{x}_0) $ suggests strong alignment between $ \boldsymbol{\omega}_{t_0}$ and $ \mathbf{e}_{02}(\boldsymbol{x}_0,t_0) $.  Specifically, this area displays high stretching and compression in the $ \mathbf{e}_{03}(\boldsymbol{x}_0,t_0) $ and $ \mathbf{e}_{01}(\boldsymbol{x}_0,t_0) $ directions (see panels c,f).  Therefore, the deformation principally occurs on a 2D-plane spanned by $ \mathbf{e}_{03}(\boldsymbol{x}_0,t_0) $ and $ \mathbf{e}_{01}(\boldsymbol{x}_0,t_0) $. The vorticity is constrained to point out of that plane and align with $ \mathbf{e}_{02}(\boldsymbol{x}_0) $. Such high strain areas have frequently been associated with regions where the vorticity preferentially aligns with the intermediate eigenvector of the rate\textendash of\textendash strain tensor \citep{Guala2005}.   Note, however,  that regions of intense alignment between the $ \boldsymbol{\omega}_{t_0} $ and $ \mathbf{e}_{02}(\boldsymbol{x}_0,t_0) $ can also arise in low strain areas (see white inset in figure \ref{fig:Fig6}).  We emphasize that in those areas,  the estimate $ \mathrm{L}_{t_0-20\tau_{\eta}}^{t_0+20\tau_{\eta}}(\boldsymbol{x}_0) $ guarantees preferential alignment between $ \boldsymbol{\omega}_{t_0} $ and $ \mathbf{e}_{02}(\boldsymbol{x}_0,t_0) $.
\\ \\
Figure \ref{fig:Fig7} displays the cumulative distributive functions (CDF) for the scalar fields from figure \ref{fig:Fig6}. For $ 65 \% $ of the initial conditions, the vorticity is preferentially aligned with $ \mathbf{e}_{02}(\boldsymbol{x}_0) $. Additionally, the CDFs in panel b show that the vorticity is more likely to be orthogonal to $ \mathbf{e}_{01}(\boldsymbol{x}_0) $ than to $ \mathbf{e}_{03}(\boldsymbol{x}_0) $. This is also reflected in the distribution of $ \mathrm{L}_{t_0}^{t_0+20\tau_{\eta}}(\boldsymbol{x}_0) $ and $ \mathrm{L}_{t_0}^{t_0-20\tau_{\eta}}(\boldsymbol{x}_0) $ (see panel a).   We attribute this to the fact that the CDF of $ \sigma_{02}(\boldsymbol{x}_0,t_0) $ is skewed towards positive values and therefore $ \sigma_{03}(\boldsymbol{x}_0,t_0) $ is on average smaller than $ -\sigma_{01}(\boldsymbol{x}_0,t_0) $.  Therefore,  the exponential decay rate of the forward integral is smaller than the one of the backward integral.  \\ \\
By taking into account the effect of viscous and non-conservative forces as well as the strain rate variation along a trajectory, the asymptotic estimate $ \mathrm{L}_{t_0-20\tau_{\eta}}^{t_0+20\tau_{\eta}} $ guarantees that for $ 15 \% $ of the initial conditions, the vorticity preferentially aligns with the intermediate Eulerian stretching direction (see panel a). Therefore, our results for homogenous isotropic turbulence show that Theorem \ref{thm: 2} provides a conservative but physically insightful upper bound on the squared sine of the angle between the vorticity and the intermediate eigenvector of the rate-of-strain tensor.
\section{Conclusions}
We have reconsidered here the phenomenon of vorticity alignment (or lack thereof) with the Lyapunov vectors and various distinguished Eulerian stretching directions in flows with pointwise bounded vorticity. As a first observation, we have pointed out that any observed alignment of vorticity with distinguished Eulerian or Lagrangian strain directions is a fundamentally frame-dependent phenomenon. This follows from the frame-dependence of the direction of the vorticity vector and the frame-indifference of the eigenvectors of common Eulerian and Lagrangian strain tensors. As a consequence, any observed preferential alignment reported so far is specific to inertial frames and will not persist in other frames. \\ \\ Next, we have derived a general Lagrangian estimate for the angle between the vorticity vector and the asymptotic limits of the eigenvectors of the Cauchy\textendash Green strain tensor, whenever those limits exist. This estimate enabled us to conclude that in inviscid flows with pointwise uniformly bounded vorticity and conservative body forces, the vorticity must be exactly orthogonal to the planes spanned by the dominant forward and backward Lyapunov vectors (Theorem \ref{thm: 1}) along trajectories with positive forward and backward maximal Lyapunov exponents. For steady and time-periodic flows defined on compact domains, this conclusion can be strengthened, given that the Lyapunov exponents are known to exist for almost all initial conditions by Oseledets multiplicative ergodic theorem (Corollary \ref{cor: 1}). 
If the forward- and backward asymptotic behaviors along a trajectory coincide, then Corollary 1 implies a perfect alignment between the vorticity and the intermediate Lyapunov vector in inviscid flows (Remark 1). To our knowledge, this is the first exact mathematical result that establishes preferential vorticity alignment with intermediate stretching directions, which is the Lagrangian analog of the empirically observed preferential alignment between the vorticity and the intermediate eigenvector of the rate-of-strain tensor in the Eulerian setting. \\ \\
Our results underline that the previously reported asymptotic alignment between the vorticity and the left singular vector of the deformation gradient cannot happen along trajectories with bounded vorticity and positive Lyapunov exponents (Remark 2). Indeed, such an alignment would imply temporally unbounded vorticity growth and hence establish a contradiction. In homogeneous isotropic turbulence, most initial conditions are expected to have positive Lyapunov exponents, and hence numerical indications of vorticity alignment with the most stretching direction are likely to arise from the unavoidable numerical errors arising in the computation of Lagrangian strain eigenvectors over long time intervals. \\ \\
Finally, we considered viscous flows with non-conservative body forces and with pointwise bounded vorticity. For such flows, our Theorem \ref{thm: 2} gives an exact upper estimate on the deviation between the vorticity and the intermediate eigenvector of the rate-of-strain tensor. The asymptotic estimate provides physical conditions such that the vorticity preferentially aligns with the intermediate eigenvector of the rate-of-strain tensor. This alignment is strong in high strain regions given that over short trajectory segments the variation of the rate-of-strain tensor is small and the curl of viscous and non-conservative forces are small relative to the initial vorticity magnitude. We have confirmed this prediction in DNS data on homogeneous and isotropic turbulence. 


\newpage
\appendix
\section{Proof of Theorem 2}
\label{app: Proof of Theorem 2}
We recall from section \ref{sec:EulerianResults} the integral form of the vorticity transport equation (\ref{eq:vorticity transport-1}),
\begin{equation}
\boldsymbol{\omega}_{t}\left(\boldsymbol{x}_0\right)=e^{\mathbf{S}_{0}(\boldsymbol{x}_0,t_{0})\left(t-t_{0}\right)}\boldsymbol{\omega}_{t_0}\left(\boldsymbol{x}_0\right)+\int_{t_{0}}^{t}e^{\mathbf{S}_{0}(\boldsymbol{x}_0,t_{0})\left(t-s\right)}\boldsymbol{g}_s(\boldsymbol{x}_0;\nu, \boldsymbol{f})\,ds, \label{eq:Appendix1}
\end{equation} where 
\begin{equation}
\boldsymbol{g}_t(\boldsymbol{x}_0;\nu, \boldsymbol{f}) :=\nu\Delta\boldsymbol{\omega}_t(\boldsymbol{x}_{0})+\boldsymbol{\nabla}\times\boldsymbol{f}(\boldsymbol{x}(t;t_0,\boldsymbol{x}_{0}),t)+\boldsymbol{\Gamma}\left(\boldsymbol{x}_0,t\right)\boldsymbol{\omega}_t(\boldsymbol{x}_{0})
\end{equation} contains the contribution of the viscous and non-potential external forces,  and the strain rate variation $ \mathbf{\Gamma} $ times the vorticity.
Squaring both sides of Eq.\ref{eq:Appendix1} yields
\begin{align}
\left| \boldsymbol{\omega}_{t}(\boldsymbol{x}_0) \right|^2& = \left| e^{\mathbf{S}_0(\boldsymbol{x}_0,t_0)(t-t_0)} \boldsymbol{\omega}_{t_0} \right|^2 \nonumber \\&+2\left\langle  e^{\mathbf{S}_0(\boldsymbol{x}_0,t_0)(t-t_0)} \boldsymbol{\omega}_{t_0}, \int_{t_{0}}^{t}e^{\mathbf{S}_{0}(\boldsymbol{x}_0,t_{0})\left(t-s\right)}\boldsymbol{g}_s(\boldsymbol{x}_0;\nu, \boldsymbol{f})\,ds\right\rangle \nonumber \\
&+ \left| \int_{t_{0}}^{t}e^{\mathbf{S}_{0}(\boldsymbol{x}_0,t_{0})\left(t-s\right)}\boldsymbol{g}_s(\boldsymbol{x}_0;\nu, \boldsymbol{f})\,ds \right|^2 \label{eq:Appendix2}
\end{align}
The eigenvalue decomposition from formula (\ref{eq:e_0j}) implies
\begin{equation} e^{\mathbf{S}_0(\boldsymbol{x}_0,t_0)t}\boldsymbol{e}_{0j}(\boldsymbol{x}_0,t_0) = e^{\sigma_{0j}(\boldsymbol{x}_0,t_0)t}\mathbf{e}_{0j}(\boldsymbol{x}_0,t_0) \label{eq: e_0j_appendix}
\end{equation} and we can therefore write
\begin{align}
&\left|e^{\mathbf{S}_0(\boldsymbol{x}_0,t_0)(t-t_0)}\boldsymbol{\omega}_{t_{0}}\left(\boldsymbol{x}_{0}\right)\right|^{2} \nonumber \\
&=\left|e^{\mathbf{S}_0(\boldsymbol{x}_0,t_0)t}\sum_{j=1}^{3}\left\langle \boldsymbol{\omega}_{t_{0}}\left(\boldsymbol{x}_{0}\right),\mathbf{e}_{0j}(\boldsymbol{x}_{0},t_0)\right \rangle \mathbf{e}_{0j}(\boldsymbol{x}_{0},t_0)\right|^{2} \nonumber \\
 &=\left|\sum_{j=1}^{3}\left\langle \boldsymbol{\omega}_{t_{0}}\left(\boldsymbol{x}_{0}\right),\mathbf{e}_{0j}(\boldsymbol{x}_{0},t_0)\right\rangle e^{\sigma_{0j}(\boldsymbol{x}_0,t_0)t}\mathbf{e}_{0j}(\boldsymbol{x}_0,t_0)\right|^{2}\nonumber \\
 &=\sum_{j=1}^{3}\left\langle \boldsymbol{\omega}_{t_{0}}\left(\boldsymbol{x}_{0}\right),\mathbf{e}_{0j}(\boldsymbol{x}_0,t_0)\right\rangle ^{2} e^{2\sigma_{0j}(\boldsymbol{x}_0,t_0)t} \nonumber \\
 &\geq\left\langle \boldsymbol{\omega}_{t_{0}}\left(\boldsymbol{x}_{0}\right),\mathbf{e}_{0j}(\boldsymbol{x}_0,t_0)\right\rangle ^{2}e^{2\sigma_{0j}(\boldsymbol{x}_0,t_0)t},\qquad j=1,2,3.\label{eq:Appendix3}
\end{align}
Combining the last inequality with Eq.(\ref{eq:Appendix2}) gives
\begin{align}
&\left\langle\boldsymbol{\omega}_{t_{0}}\left(\boldsymbol{x}_{0}\right),\mathbf{e}_{03}(\boldsymbol{x}_0,t_0)\right\rangle^{2}e^{2\sigma_{03}(\boldsymbol{x}_0,t_0)t} \nonumber \\ 
&\leq\left| \boldsymbol{\omega}_{t}(\boldsymbol{x}_0) \right|^2-2\left\langle  e^{\mathbf{S}_0(\boldsymbol{x}_0,t_0)(t-t_0)} \boldsymbol{\omega}_{t_0}, \int_{t_{0}}^{t}e^{\mathbf{S}_{0}(\boldsymbol{x}_0,t_{0})\left(t-s\right)}\boldsymbol{g}_s(\boldsymbol{x}_0;\nu, \boldsymbol{f})\,ds\right\rangle \nonumber \\
&-\left| \int_{t_{0}}^{t}e^{\mathbf{S}_{0}(\boldsymbol{x}_0,t_{0})\left(t-s\right)}\boldsymbol{g}_s(\boldsymbol{x}_0;\nu, \boldsymbol{f})\,ds \right|^2 \nonumber \\
&\leq  \left| \boldsymbol{\omega}_{t}(\boldsymbol{x}_0) \right|^2+2\left|\boldsymbol{\omega}_{t_0}\right| \left| \int_{t_{0}}^{t}e^{\mathbf{S}_{0}(\boldsymbol{x}_0,t_{0})\left(2t-t_0-s\right)}\boldsymbol{g}_s(\boldsymbol{x}_0;\nu, \boldsymbol{f})\,ds\right| \nonumber \\
&- \left| \int_{t_{0}}^{t}e^{\mathbf{S}_{0}(\boldsymbol{x}_0,t_{0})\left(t-s\right)}\boldsymbol{g}_s(\boldsymbol{x}_0;\nu, \boldsymbol{f})\,ds \right|^2.  \label{eq: Appendix4}
\end{align}
By dividing both sides of Eq.(\ref{eq: Appendix4}) by $ e^{2 \sigma_{03}(\boldsymbol{x}_0,t_0)(t-t_0)} $ and introducing the notation \[ \mathbf{e}_{\omega_{t_0}}(\boldsymbol{x}_0) := \frac{\boldsymbol{\omega}_{t_0}(\boldsymbol{x}_0)}{\left| \boldsymbol{\omega}_{t_0}(\boldsymbol{x}_0) \right|}, \] we re-write inequality (\ref{eq: Appendix4}) as
\begin{align}
&\left\langle\mathbf{e}_{\omega_{t_0}},\mathbf{e}_{03}(\boldsymbol{x}_0,t_0)\right\rangle^{2} \nonumber \\ 
&\leq \frac{\left|   \boldsymbol{\omega}_{t}(\boldsymbol{x}_0)\right|^2}{\left|   \boldsymbol{\omega}_{t_0}(\boldsymbol{x}_0)\right|^2}e^{-2 \sigma_{03}(\boldsymbol{x}_0,t_0)(t-t_0)}+2\frac{\left|\displaystyle\int_{t_0}^t e^{\mathbf{S}_0(\boldsymbol{x}_0,t_0)(2t-t_0-s)} \boldsymbol{g}_s(\boldsymbol{x}_0;\nu, \boldsymbol{f})\,ds  \right|}{ \left| \boldsymbol{\omega}_{t_0}(\boldsymbol{x}_0,t_0) \right|e^{2\sigma_{03}(\boldsymbol{x}_0,t_0)(t-t_0)}} \nonumber \\
&-\left(\frac{\left|\displaystyle\int_{t_0}^t e^{\mathbf{S}_0(\boldsymbol{x}_0,t_0)(t-s)} \boldsymbol{g}_s(\boldsymbol{x}_0;\nu, \boldsymbol{f})\,ds  \right|}{ \left| \boldsymbol{\omega}_{t_0}(\boldsymbol{x}_0,t_0) \right|e^{\sigma_{03}(\boldsymbol{x}_0,t_0)(t-t_0)}}\right)^2 \nonumber \\
&= e^{2\left(\mathrm{VSE}_{t_0}^t(\boldsymbol{x}_0)- \sigma_{03}(\boldsymbol{x}_0,t_0)\right)(t-t_0)}+2\frac{\left|\displaystyle\int_{t_0}^t e^{\mathbf{S}_0(\boldsymbol{x}_0,t_0)(2t-t_0-s)} \boldsymbol{g}_s(\boldsymbol{x}_0;\nu, \boldsymbol{f})\,ds  \right|}{ \left| \boldsymbol{\omega}_{t_0}(\boldsymbol{x}_0,t_0) \right|e^{2\sigma_{03}(\boldsymbol{x}_0,t_0)(t-t_0)}} \nonumber \\
&-\left(\frac{\left|\displaystyle\int_{t_0}^t e^{\mathbf{S}_0(\boldsymbol{x}_0,t_0)(t-s)} \boldsymbol{g}_s(\boldsymbol{x}_0;\nu, \boldsymbol{f})\,ds  \right|}{ \left| \boldsymbol{\omega}_{t_0}(\boldsymbol{x}_0,t_0) \right|e^{\sigma_{03}(\boldsymbol{x}_0,t_0)(t-t_0)}}\right)^2, \label{eq: Appendix5}
\end{align} where $ \mathrm{VSE}_{t_0}^t(\boldsymbol{x}_0) $ is the vorticity stretching exponent defined in formula (\ref{eq:VSEdef}).  For flows with  bounded vorticity the first term in the inequality (\ref{eq: Appendix5}) vanishes in the asymptotic limit $  t \rightarrow \infty $.
By using the eigenvalue decomposition (\ref{eq: e_0j_appendix}), we write out the individual terms in Eq.(\ref{eq: Appendix5}) as
\begin{align}
\label{eq: Inequality_Eigenvalue_Decomposition_1}
&\frac{\left|\displaystyle\int_{t_0}^t e^{\mathbf{S}_0(\boldsymbol{x}_0,t_0)(2t-t_0-s)} \boldsymbol{g}_s(\boldsymbol{x}_0;\nu, \boldsymbol{f})\,ds  \right|}{e^{2\sigma_{03}(\boldsymbol{x}_0,t_0)(t-t_0)}}\\
&= \left| \renewcommand\arraystretch{3.5}\begin{pmatrix} \displaystyle\int_{t_0}^t\frac{e^{\sigma_{01}(\boldsymbol{x}_0,t_0)(2t-t_0-s)}}{e^{2\sigma_{03}(\boldsymbol{x}_0,t_0)(t-t_0)}} \left\langle \mathbf{e}_{01}(\boldsymbol{x}_0,t_0), \boldsymbol{g}_s(\boldsymbol{x}_0;\nu, \boldsymbol{f})\right\rangle ds \nonumber \\
\displaystyle\int_{t_0}^t\frac{e^{\sigma_{02}(\boldsymbol{x}_0,t_0)(2t-t_0-s)}}{e^{2\sigma_{03}(\boldsymbol{x}_0,t_0)(t-t_0)}} \left\langle \mathbf{e}_{02}(\boldsymbol{x}_0,t_0), \boldsymbol{g}_s(\boldsymbol{x}_0;\nu, \boldsymbol{f}) \right\rangle ds \nonumber \\
\displaystyle\int_{t_0}^te^{\sigma_{03}(\boldsymbol{x}_0,t_0)(t_0-s)} \left\langle \mathbf{e}_{03}(\boldsymbol{x}_0,t_0), \boldsymbol{g}_s(\boldsymbol{x}_0;\nu, \boldsymbol{f})\right\rangle ds \end{pmatrix} \right| \nonumber,
\end{align}
and 
\begin{align}
\label{eq: Inequality_Eigenvalue_Decomposition_2}
&\frac{\left|\displaystyle\int_{t_0}^t e^{\mathbf{S}_0(\boldsymbol{x}_0,t_0)(t-s)} \boldsymbol{g}_s(\boldsymbol{x}_0;\nu, \boldsymbol{f})\,ds  \right|}{e^{\sigma_{03}(\boldsymbol{x}_0,t_0)(t-t_0)}} \\
&= \left| \renewcommand\arraystretch{3.5}\begin{pmatrix} \displaystyle\int_{t_0}^t\frac{e^{\sigma_{01}(\boldsymbol{x}_0,t_0)(t-s)}}{e^{\sigma_{03}(\boldsymbol{x}_0,t_0)(t-t_0)}} \left\langle \mathbf{e}_{01}(\boldsymbol{x}_0,t_0), \boldsymbol{g}_s(\boldsymbol{x}_0;\nu, \boldsymbol{f})\right\rangle ds \\
\displaystyle\int_{t_0}^t\frac{e^{\sigma_{02}(\boldsymbol{x}_0,t_0)(t-s)}}{e^{\sigma_{03}(\boldsymbol{x}_0,t_0)(t-t_0)}} \left\langle \mathbf{e}_{02}(\boldsymbol{x}_0,t_0), \boldsymbol{g}_s(\boldsymbol{x}_0;\nu, \boldsymbol{f}) \right\rangle ds \\
\displaystyle\int_{t_0}^te^{\sigma_{03}(\boldsymbol{x}_0,t_0)(t_0-s)} \left\langle \mathbf{e}_{03}(\boldsymbol{x}_0,t_0), \boldsymbol{g}_s(\boldsymbol{x}_0;\nu, \boldsymbol{f})\right\rangle ds \end{pmatrix} \right|.
\end{align}
With the notation
\[ \left\Vert \boldsymbol{g} \right\Vert_{\infty} = \sup\limits_{t \in \mathbb{R}, \boldsymbol{x}_0 \in U} \left| \boldsymbol{g}_t(\boldsymbol{x}_0,t_0;\nu,\boldsymbol{f}) \right|,\] we then have for the individual terms in formulas (\ref{eq: Inequality_Eigenvalue_Decomposition_1})-(\ref{eq: Inequality_Eigenvalue_Decomposition_2})
\begin{align}
\label{eq: upperbound_Inequality1}
&\displaystyle\int_{t_0}^t \frac{e^{\sigma_{01}(\boldsymbol{x}_0,t_0)(2t-t_0-s)}}{e^{2\sigma_{03}(\boldsymbol{x}_0,t_0)(t-t_0)}}\left\langle \mathbf{e}_{01}(\boldsymbol{x}_0,t_0), \boldsymbol{g}_s(\boldsymbol{x}_0;\nu, \boldsymbol{f})\right\rangle ds \\
&\leq \frac{e^{\sigma_{01}(\boldsymbol{x}_0,t_0)(2t-t_0)}}{e^{2\sigma_{03}(\boldsymbol{x}_0,t_0)(t-t_0)}} \displaystyle\int_{t_0}^te^{-\sigma_{01}(\boldsymbol{x}_0,t_0)s}ds \left\Vert \boldsymbol{g} \right\Vert_{\infty} \nonumber \\
&= \frac{e^{\sigma_{01}(\boldsymbol{x}_0,t_0)(2t-t_0)}}{e^{2\sigma_{03}(\boldsymbol{x}_0,t_0)(t-t_0)}}\left[\frac{e^{-\sigma_{01}(\boldsymbol{x}_0,t_0)t_0}-e^{-\sigma_{01}(\boldsymbol{x}_0,t_0)t}}{\sigma_{01}(\boldsymbol{x}_0,t_0)}\right]\left\Vert \boldsymbol{g} \right\Vert_{\infty} \nonumber\\
&= \left[\frac{e^{2\sigma_{01}(\boldsymbol{x}_0,t_0)(t-t_0)}-e^{\sigma_{01}(\boldsymbol{x}_0,t_0)(t-t_0)}}{e^{2\sigma_{03}(\boldsymbol{x}_0,t_0)(t-t_0)}}\right]\frac{\left\Vert \boldsymbol{g} \right\Vert_{\infty}}{\sigma_{01}(\boldsymbol{x}_0,t_0)} \nonumber,
\end{align} and similarly
\begin{align}
&\int_{t_0}^t \frac{e^{\sigma_{02}(\boldsymbol{x}_0,t_0)(2t-t_0-s)}}{e^{2\sigma_{03}(\boldsymbol{x}_0,t_0)(t-t_0)}}\left\langle \mathbf{e}_{02}(\boldsymbol{x}_0,t_0), \boldsymbol{g}_s(\boldsymbol{x}_0;\nu, \boldsymbol{f})\right\rangle ds \nonumber \\
&\leq \left[\frac{e^{2\sigma_{02}(\boldsymbol{x}_0,t_0)(t-t_0)}-e^{\sigma_{02}(\boldsymbol{x}_0,t_0)(t-t_0)}}{e^{2\sigma_{03}(\boldsymbol{x}_0,t_0)(t-t_0)}}\right]\frac{\left\Vert\boldsymbol{g}\right\Vert_{\infty}}{\sigma_{02}(\boldsymbol{x}_0,t_0)},\label{eq: upperbound_Inequality2} \\
&\int_{t_0}^t \frac{e^{\sigma_{01}(\boldsymbol{x}_0,t_0)(t-s)}}{e^{\sigma_{03}(\boldsymbol{x}_0,t_0)(t-t_0)}}\left\langle \mathbf{e}_{01}(\boldsymbol{x}_0,t_0), \boldsymbol{g}_s(\boldsymbol{x}_0;\nu, \boldsymbol{f})\right\rangle ds \nonumber \\ 
&\leq e^{-\sigma_{03}(\boldsymbol{x}_0,t_0)(t-t_0)}\left[1-e^{\sigma_{01}(\boldsymbol{x}_0,t_0)(t-t_0)}\right]\frac{\left\Vert\boldsymbol{g}\right\Vert_{\infty}}{\sigma_{01}(\boldsymbol{x}_0,t_0)},\label{eq: upperbound_Inequality3} \\
&\int_{t_0}^t \frac{e^{\sigma_{02}(\boldsymbol{x}_0,t_0)(t-s)}}{e^{\sigma_{03}(\boldsymbol{x}_0,t_0)(t-t_0)}}\left\langle \mathbf{e}_{02}(\boldsymbol{x}_0,t_0), \boldsymbol{g}_s(\boldsymbol{x}_0;\nu, \boldsymbol{f})\right\rangle ds \nonumber \\ 
&\leq e^{-\sigma_{03}(\boldsymbol{x}_0,t_0)(t-t_0)}\left[1-e^{\sigma_{02}(\boldsymbol{x}_0,t_0)(t-t_0)}\right]\frac{\left\Vert\boldsymbol{g}\right\Vert_{\infty}}{\sigma_{02}(\boldsymbol{x}_0,t_0)},\label{eq: upperbound_Inequality4}
\end{align}
Except for trajectories where the eigenvalues of $ \mathbf{S}_0(\boldsymbol{x}_0, t_0) $ are repeated, i.e., at singularities of $ \mathbf{S}_0(\boldsymbol{x}_0,t_0) $, the right hand side of the inequalities (\ref{eq: upperbound_Inequality1})-(\ref{eq: upperbound_Inequality4}) vanishes in the asymptotic limit $  t \rightarrow \infty $.  By combining the latter statements, we obtain the following simplified version of inequality (\ref{eq: Appendix5}) in the asymptotic limit $  t \rightarrow \infty $:
\begin{align}
&\left\langle\mathbf{e}_{\omega_{t_0}},\mathbf{e}_{03}(\boldsymbol{x}_0,t_0)\right\rangle^{2} \nonumber \\
&\leq 2\frac{\left|\displaystyle\int_{t_0}^{\infty}e^{\sigma_{03}(\boldsymbol{x}_0,t_0)(t_0-s)} \left\langle \mathbf{e}_{03}(\boldsymbol{x}_0,t_0), \boldsymbol{g}_s(\boldsymbol{x}_0;\nu, \boldsymbol{f})\right\rangle ds  \right|}{ \left| \boldsymbol{\omega}_{t_0}(\boldsymbol{x}_0,t_0) \right|} \nonumber\\
&-\left(\frac{\left|\displaystyle\int_{t_0}^{\infty}e^{\sigma_{03}(\boldsymbol{x}_0,t_0)(t_0-s)} \left\langle \mathbf{e}_{03}(\boldsymbol{x}_0,t_0), \boldsymbol{g}_s(\boldsymbol{x}_0;\nu, \boldsymbol{f})\right\rangle ds \right|}{ \left| \boldsymbol{\omega}_{t_0}(\boldsymbol{x}_0,t_0) \right|}\right)^2 \nonumber\\
&= 1-\left(1 -\frac{\left|\displaystyle\int_{t_0}^{\infty}e^{\sigma_{03}(\boldsymbol{x}_0,t_0)(t_0-s)} \left\langle \mathbf{e}_{03}(\boldsymbol{x}_0,t_0), \boldsymbol{g}_s(\boldsymbol{x}_0;\nu, \boldsymbol{f})\right\rangle ds  \right|}{ \left| \boldsymbol{\omega}_{t_0}(\boldsymbol{x}_0,t_0) \right|}\right)^2 \nonumber \\
&=  1-\left(1 -\left|\displaystyle\int_{t_0}^{\infty}e^{\sigma_{03}(\boldsymbol{x}_0,t_0)(t_0-s)} \left\langle \mathbf{e}_{03}(\boldsymbol{x}_0,t_0), \boldsymbol{h}_s(\boldsymbol{x}_0;\nu, \boldsymbol{f})\right\rangle ds  \right|\right)^2 \label{eq: Appendix6},
\end{align}
where we introduced the notation
\begin{align*}
\boldsymbol{h}_t(\boldsymbol{x}_0;\nu, \boldsymbol{f}) &:= \frac{\nu\Delta\boldsymbol{\omega}_t(\boldsymbol{x}_{0})+\boldsymbol{\nabla}\times\boldsymbol{f}_t(\boldsymbol{x}_{0})+\boldsymbol{\Gamma}\left(\boldsymbol{x}_0,t\right)\boldsymbol{\omega}_t(\boldsymbol{x}_{0})}{\left| \boldsymbol{\omega}_{t_0}(\boldsymbol{x}_0,t_0) \right|}\\
&=\frac{\boldsymbol{g}_t(\boldsymbol{x}_0;\nu, \boldsymbol{f})}{\left| \boldsymbol{\omega}_{t_0}(\boldsymbol{x}_0,t_0) \right|}.
\end{align*}
Applying the same reasoning in backward time and taking the asymptotic limit $ t \rightarrow -\infty $ yields
\begin{align}
&\left\langle\mathbf{e}_{\omega_{t_0}},\mathbf{e}_{01}(\boldsymbol{x}_0,t_0)\right\rangle^{2} \leq 1- \left(1 - \right. \nonumber \\
& \left. \left|\displaystyle\int_{t_0}^{-\infty}e^{\sigma_{01}(\boldsymbol{x}_0,t_0)(t_0-s)} \left\langle \mathbf{e}_{01}(\boldsymbol{x}_0,t_0), \boldsymbol{h}_s(\boldsymbol{x}_0;\nu, \boldsymbol{f})\right\rangle ds  \right|\right)^2 \label{eq: Appendix7}
\end{align}
The eigenvectors $ \mathbf{e}_{0j}(\boldsymbol{x}_0,t_0) $ are mutually orthogonal and hence
\begin{equation}
\left\langle \mathbf{e}_{01}(\boldsymbol{x}_0,t_0),  \mathbf{e}_{\omega_{t_0}}\right\rangle^2+\left\langle \mathbf{e}_{02}(\boldsymbol{x}_0,t_0),  \mathbf{e}_{\omega_{t_0}}\right\rangle^2+\left\langle \mathbf{e}_{03}(\boldsymbol{x}_0,t_0),  \mathbf{e}_{\omega_{t_0}}\right\rangle^2 = 1.
\end{equation} The angle $ \beta(\boldsymbol{x}_0,t_0) $ between the vorticity $ \boldsymbol{\omega}_0 $ and the intermediate eigenvector $ \mathbf{e}_{02}(\boldsymbol{x}_0,t_0) $ of the rate-of-strain tensor then satisfies
\begin{align}
\sin^2\beta(\boldsymbol{x}_0,t_0)=\left\langle \mathbf{e}_{01}(\boldsymbol{x}_0,t_0),  \mathbf{e}_{\omega_{t_0}}\right\rangle^2+\left\langle \mathbf{e}_{03}(\boldsymbol{x}_0,t_0),  \mathbf{e}_{\omega_{t_0}}\right\rangle^2. \label{eq: Appendix8}
\end{align}
By combining the last equation with the inequalities (\ref{eq: Appendix6})-(\ref{eq: Appendix7}) we obtain
\begin{align}
&\sin^2\beta(\boldsymbol{x}_0,t_0) \leq 2 - \nonumber \\
&\left(1-\left|\displaystyle\int_{t_0}^{\infty}e^{\sigma_{03}(\boldsymbol{x}_0,t_0)(t_0-s)} \left\langle \mathbf{e}_{03}(\boldsymbol{x}_0,t_0), \boldsymbol{h}_s(\boldsymbol{x}_0;\nu, \boldsymbol{f})\right\rangle ds  \right|\right)^2 - \nonumber \\
&\left(1 -\left|\displaystyle\int_{t_0}^{-\infty}e^{\sigma_{01}(\boldsymbol{x}_0,t_0)(t_0-s)} \left\langle \mathbf{e}_{01}(\boldsymbol{x}_0,t_0), \boldsymbol{h}_s(\boldsymbol{x}_0;\nu, \boldsymbol{f})\right\rangle ds  \right|\right)^2 \label{eq: Appendix9}
\end{align}
This concludes the proof of Theorem \ref{thm: 2}.



\bibliographystyle{elsarticle-num-names} 
\bibliography{references}





\end{document}